# Observation of Integer and Fractional Chern insulators in high Chern number flatbands

**Authors:** Jingwei Dong[1,2], Le Liu[1,2], Jundong Zhu[1,2], Zitian Pan[1,2], Yu Hong[1,2], Fankun Wang[1,2], Zhiyuan Ren[1,2], Zhengnan Jia[1,2], Kenji Watanabe[3], Takashi Taniguchi[4], Luojun Du[1,2], Dongxia Shi[1,2], Wei Yang[1,2]*, Guangyu Zhang[1,2,5]*

[1]Beijing National Laboratory for Condensed Matter Physics and Institute of Physics, Chinese Academy of Sciences, Beijing 100190, China.

[2]School of Physical Sciences, University of Chinese Academy of Sciences, Beijing, 100190, China.

[3]Research Center for Electronic and Optical Materials, National Institute for Materials Science, 1-1 Namiki, Tsukuba 305-0044, Japan.

[4]Research Center for Materials Nanoarchitectonics, National Institute for Materials Science, 1-1 Namiki, Tsukuba 305-0044, Japan

[5]Songshan Lake Materials Laboratory, Dongguan 523808, China

*Corresponding author. Email: wei.yang@iphy.ac.cn; gyzhang@iphy.ac.cn

**Abstract:** Moire flatbands with high Chern numbers (C>1) offer opportunities to study the fractional quantum anomalous Hall effects that go beyond the Landau level paradigm with C=1, which remain unexplored yet. Here, we target the novel topological phases in high Chern number flatbands by designing a new moiré system, i.e., twisted rhombohedral trilayer-bilayer graphene. We observe quantized anomalous Hall effects (QAH) with $C = 3$ at $\nu = 1$ and $\nu = 3$, demonstrating the high Chern number nature of the flat band from continuum calculations. By fractionally filling the flat band, we observe QAH with $C = 2$ at even-denominator fractional filling $\nu = 3/2$, as well as QAH with $C = 3$ continuously from $\nu = 1$ to the even-denominator $\nu = 3/2$ at zero magnetic fields. Most importantly, we observe, for the first time, evidence of an FCI with $C = -6/5$ at v = 12/5, corresponding to 2/5 filling of a high Chern number flat band with C = -3, verified by both Streda's formula analysis and a fractionally QAH. It is also worth noting that Streda's formula analysis reveals a signature of another FCI with $C = -3/2$ at $\nu = 5/2$ under finite magnetic fields. Our results demonstrate the tRTBG, which can be naturally extended to other twisted graphene moiré superlattices based on rhombohedral graphene multilayers, as a novel platform for hosting unconventional high Chern number correlated topology in the ultra-strong correlated regime that is beyond the paradigm of fractional phases with $C < 1$.

**Main Text:** The Chern insulator (*1*), characterized by a topological Chern number (*C*) defined by quantum geometry of the bands, is a canonical result of the intriguing interplay between correlation and topology. Being an analog of the quantum Hall insulator at zero magnetic fields, Chern insulator is famous for producing quantized anomalous Hall effect (*2*) with $R_{xy} = h/Ce^2$ that has been observed in many different systems, including magnetically doped topological insulators $(Bi,Sb)_2Te_3$ (*3*) and $MnBi_2Te_4$ (*4*), and graphene and TMDC based multilayers(*5-15*). More intriguingly, fractional Chern insualtors (*16-20*)’, an analog of the fractional quantum Hall insulator (*20-22*) at zero magnetic fields, have been observed in the fractional quantum anomalous Hall effect (fQAHE) in pioneering works of two different moiré systems, i.e. twisted $MoTe_2$ (*23-26*) and rhombohedral pentalayer graphene/hBN moiré superlattices (*10, 13*). The fractional excitations (*27, 28*), including non-Abelian anyons (*29*), are believed to hold great promise for the next generation of electronics and quantum computing (*30*). Thus far, fQAHE are observed by fractionally filling moiré flatband with $C = 1$. Exploration of new fractional phases in high Chern number ($C > 1$) flatband could even go beyond the fQHE analog (*31, 32*) and is yet to be discovered.

Of particular interest is the fQAHE in graphene multilayers(*10*), where both the rhombohedral stacking order and the moiré superlattice are important. The rhombohedral stacking order gives low energy band dispersion that can be approximated as $E \sim k^N$ (N is the layer number) in a single particle picture (*33*), and it eventually leads to a narrow flat band condition by applying displacement fields. The moiré superlattice further flattens the bands due to the reduced size of the moiré Brillouin zone, favoring time-reversal symmetry breaking in the flat band limit (*34, 35*), and brings its Berry curvature (*36-38*) to meet the criterion for the emergence of fQAHE. Recently, extended QAHE with $C = 1$, developing from moiré filling $\nu = 0.5$ to 1.3, also named as anomalous Hall crystal, is observed competing against fQAHE in rhombohedral pentalayer graphene aligned with hBN (*39*). In addition, graphene multilayers could host topological flatbands with high Chern numbers (*40-43*). Topological bands with high Chern numbers $C = 4$ and 5 are observed in rhombohedral graphene tetralayer and pentalayer (*11, 12*), respectively.

Here, we target novel topological phases with high Chern numbers by designing a new rhombohedral graphene moiré superlattice, i.e. twisted rhombohedral trilayer-bilayer graphene (tRTBG) moiré system. Compared to the rhombohedral multilayer graphene/hBN moiré superlattice, tRTBG has two advantages, i.e. stronger moiré potential from the moiré between graphene multilayers, and high Chern numbers from rhombohedral stacking order (*11, 12*). In this work, we fabricate tRTBG devices with twist angles from 1.2 to 1.5 degrees. And importantly, tRTBG, calculated from the continuum model, shows an isolated flat moiré conduction band and a high valley Chern number of $C_v = 3$.

**Device structure and band structure**

Figure 1A shows the schematics of the dual gate device with tRTBG moiré structure encapsulated by hBN. The rhombohedral graphene multilayers are confirmed by infrared imaging and Raman spectrum (Fig. S1). In this study, we mainly focus on three devices with twist angles ranging from 1.3 to 1.5 degrees, whose experimental parameters are summarized in Table 1. Fig.1B shows a phase diagram $R_{xx}$ ($v$, $D$) at $T$ = 1 K for device D1 with $\theta$ = 1.35° (twist angle $\theta$, moiré filling $v$, and displacement field $D$ are obtained by following the Method section). Energy gaps are observed at CNP and full filling with $v$ = 4 for $D$ > 0.2 V/nm, indicating an isolated moiré conduction band condition. Note that $D$ is positive when the field points from the bilayer graphene (bottom) to the trilayer (top), as indicated by the black arrow in Fig. 1A. Importantly, correlated insulators are observed at integer fillings $v$ = 1, 2, and 3 for positive $D$ > 0.2 V/nm (Fig. 1B). In addition, evidence of nontrivial Hall responses is observed near odd integer fillings at $v$ = 1and 3, indicated by the red and blue colors in Fig. 1C; by contrast, the correlated insulator at $v$ = 2 is topological trivial without anomalous Hall response.

The phase diagram agrees with the band structure of tRTBG calculated from the continuum model, which shows an isolated flat moiré conduction with bandwidth of ~ 6 meV and a valley Chern number of $C_v$ = 3 for $U$ = -12.5 meV (corresponding to the positive $D$, as shown in Fig. 1D). In the isolated flat band limit, strong Coulomb interactions could lead to spin and valley splitting of the four-fold degenerate moiré conduction band. Based on the nontrivial topological phases near odd fillings, spin and valley polarized subbands are developed sequentially at integer fillings, as shown schematically in Fig. 1E and 1F.

**Quantum anomalous Hall effects with *C* = 3 at *v* = 1**

To begin with, we first demonstrate the Chern insulator with $C$ = 3 at $v$ = 1. Under a given magnetic field, Chern insulators can be mapped out from longitudinal resistance minima (Fig. 2A) and the corresponding Hall resistance plateau (Fig. 2B). Fig. 2C shows typical plots of $R_{xx}$ (red) and $R_{xy}$ (blue) as a function of $v$, taken along the dashed line across $v$ = 1 in Fig. 2B. Strikingly, it shows a vanishing longitudinal resistance minimum and a corresponding quantized Hall resistance of $R_{xy} = h/(3e^2)$, indicating a Chern insulator with $C$ = 3 at $v$ = 1. We double-check the topological nature by performing Streda's formula analysis in the Landau fan diagram. As shown in Fig. 2D, the Chern insulator at $v$ = 1, with the quantized plateau $R_{xy} = h/3e^2$ together with the corresponding $R_{xx}$ minimum, onsets at zero magnetic fields and disperses with $C$ = 3 upon the increase of magnetic field B, according to $C = (h/e)dn/dB$ in Streda's formula. Smoking gun evidence of the Chern insulator is also vividly demonstrated in the quantized anomalous Hall effects in Fig. 2E, where Hall resistance is quantized with $C$ = ±3 and shows hysteresis with a steep sign switching at critical magnetic fields around ±120mT.

The observation of the Chern insulator and QAH with $C$ = 3 at $v$ = 1 clearly demonstrates the high Chern number nature of the flat band in tRTBG. The high Chern number $C$ = 3 agrees well with the orbital (valley Chern number from the band topology of tRTBG in the continuum model (Fig. 1C), which further demonstrates the rhombohedral stacking in our devices. Note that the Chern insulator, due to the valley Chern number, is also termed the orbital Chern insulator (*8, 44, 45*). It is also worth noting that QAH with $C$ = 3 at $v$ = 1 is robust with a quantization temperature up to $T$ = 3K, as well as a high critical temperature of 6K for the onset of ferromagnetism (Fig. S3). The critical temperature is comparable with respect to twisted $MoTe_2$ (*24, 26*), and about two times larger than that of Chern insulator with $C = 1$ at $\nu = 1$ in rhombohedral graphene multilayers/hBN moire superlattices (*10, 13*).

**Unconventional QAH with $C = 2$ at even-denominator fractional filling $\nu = 3/2$**

Next, we focus on the region where the moiré subbands are partially filled at $\nu = 3/2$. As shown in Fig. 2A and 2B, a signature of Chern insulator with $R_{xy}$ plateau and $R_{xx}$ minimum is revealed at even-denominator fractional filling $\nu = 3/2$ within a finite narrow range of D between 0.5 V/nm and 0.55 V/nm. To pin down the topological nature at $\nu = 3/2$, we measure the Landau fan (Fig. 2F) to carry out Streda's formula analysis. We observe a Chern insulator that onsets at zero magnetic fields and disperses with $C = 2$ at finite magnetic fields. And importantly, we also observe QAH with $C = 2$ in Fig. 2G. Compared to the Chern insulator at $\nu = 1$, QAH at $\nu = 3/2$ is more fragile with a reduced critical magnetic field of 10mT. The ferromagnetic hysteresis is very sensitive and is barely visible above 1K.

Note that both QAH with $C = 3$ at $\nu = 1$ and QAH with $C = 2$ at $\nu = 3/2$ have been reproduced in a separate high-quality device (D3) with a twist angle of 1.49°. In addition, signatures of Chern insulators are also observed at even-denominator fractional fillings $\nu = 1/2$ and 3/2 in two other devices, whose Chern numbers from Streda's law are $C = 1$ and 2, respectively (Figs. S9 and S10). Note that Chern insulator with $C = 1$ is observed at $\nu = 3/2$ in twisted monolayer-bilayer graphene (*46*).

The observation of QAH with $C = 2$ at $\nu = 3/2$, as well as the Chern insulator with $C = 1$ at $\nu = 1/2$, is unconventional. Intuitively, one might expect fractionally quantized anomalous Hall effects with a high Chern number of 3/2 at even-denominator fraction $\nu = 1/2$. The obtained $C = 2$ at $\nu = 3/2$, instead of $C = 3/2$, might indicate a non-uniform distribution of Berry curvature in tRTBG. By taking the Chern insulator with $C = 3$ as the parent state, the topological phase with $C = 1$ at $\nu = 1/2$ carries a fractional topological number of 1/3 from the high Chern number parent state, i.e. $C_{1/2} = \alpha * C_K$, with $\alpha=1/3$ and $C_K = 3$. Similarly, the Chern number for QAH at $\nu = 3/2$ is the sum of the Chern number of the first filled subband with spin and valley polarization and the fractional Chern number of the second half-filled subband with the same spin polarization yet opposite valley polarization, i.e., $C_{3/2} = C_K+\alpha * C_{-K} =3+1/3*(-3) = 2$. The same fractional coefficient α indicates a similar quantum wave function for the two symmetry-breaking subbands. Such a sum rule predicts a trivial correlated insulator with $C = 0$ at $\nu = 2$, which also agrees with our experimental observation.

Anyway, regardless of the above argument of fractional topology, the presence of integer Chern insulators at $\nu = 1/2$ and 3/2, in which both time reversal symmetry and translation symmetry are broken, indicates a coexistence of topology and charge order. They are also referred to as quantum anomalous Hall crystals (*47*), which might compete against FCI in topological moiré flat bands (*48, 49*).

**QAH with $C$ =3 developed continuously from $\nu = 1$ to $\nu = 3/2$**

In addition, QAH with $C = 3$ is observed developing continuously from $\nu = 1$ (with $D =$

0.42V/nm) to $v = 3/2$ (with $D = 0.48$V/nm), bridging the QAH with $C = 3$ at $v = 1$ and the QAH with $C = 2$ at $v = 3/2$. Fig. 2H shows Landau fan diagrams along the dashed lines in Fig. 2B. The Hall resistance is quantized over a wide range of fillings from v = 1 to 1.4, and it obeys Streda's formula with $C = 3$ (white dashed line in Fig. 2H). Fig. 2I shows one typical QAH with $C = 3$ at fractional filling $v = 1.2$.

The abovementioned observations signify the formation of quantum anomalous Hall crystals (QAHC) with a high Chern number (*50, 51*). QAHC is a topologically nontrivial analog of the Wigner crystal in the strongly correlated electrons (*48, 52-56*). It considers a system with a coexisting (anomalous) quantized Hall effect with topological properties and electron solids (also named electron crystal or Wigner crystal) with ultra-strong electron repulsion. While QHE has been widely observed and well understood, electron solids are rarely observed. In particular, the latter features the formation of a lattice-like structure with an electron density that can be tuned continuously. Recent experiments suggest that AHC can compete against fQAHE in rhombohedral graphene multilayer (*39*) and twisted $MoTe_2$ (*57*). It is also worth noting that the parent state at $v = 1$ is surrounded by a halo-like structure at the phase boundary, similar to that of TDBG (*58*). The halo structure is believed to be the van Hove singularity; in this sense, the Chern insulator at $v = 1/2$ and the AHC are also intimately tied to the VHS in the moiré flat bands. Note that we primarily focus on the experimental facts, reserving the theoretical aspects for future studies.

**QAH with $C = -3$ at $v = 3$**

Next, we focus on the nontrivial topology near $v = 3$. Fig. 3A shows a color mapping of the Hall resistance $R_{xy}(v, D)$ at $B = 0.1$T for device D2 with a twist angle of 1.5°, and it reveals a Hall resistance plateau near $v = 3$, while the corresponding $R_{xx}$ near $v = 3$ shows minima in Fig. 3B. The Landau fan diagram verifies a Chern insulator with high Chern number $C = -3$, onset at zero magnetic fields, according to the Streda formula (Fig. 2C and 2D, taken at $D = 0.384$V/nm). And importantly, we observe a quantized anomalous Hall effect at $R_{xy} = h/(3e^2)$ with a coercive field of 160 mT in Fig.3E, demonstrating the nature of high Chern number $C = 3$ topology in tRTBG. The quantized Hall resistance plateau at $v = 3$ is observed in a wide range of displacement fields from 0.35 to 0.46 V/nm (Fig. 3F). The QAH with $C = -3$ at $v = 3$ is also observed at zero magnetic fields in device D1, while there is a competing Chern insulator with $C = 3$ at high magnetic fields (Fig. S5).

**FCI with $C = -6/5$ at $v = 12/5$**

We observe the signature of fractional Chern insulators from Streda's formula analysis in device D2. Fig. 3A reveals additional topological states at fractional fillings $v = \sim 12/5$ and ~5/2 under $D = 0.36$ to 0.4 V/nm, with the Hall resistance larger than $h/3e^2$ (Fig. S6G and S6H). These topological states are dispersive with $B$ and $v$ (Fig. 3A), yielding fractional Chern numbers of $C = -6/5$ and -3/2 at $v = \sim 12/5$ and ~5/2, respectively, according to the Streda's formula analysis (red open circles and blue open triangles in Fig. 3D). The observation of fractional Chern number at fractional fillings suggests the emergence of two fractional Chern insulators (FCI) when partially filled the third high Chern number subband. $FCI_1$ with $C$ = -6/5 at $\nu$ = 12/5 onsets at zero magnetic fields, while $FCI_2$ with $C$ = -3/2 at $\nu$ = 5/2 can be stabilized by finite magnetic fields (Fig. S6I). Note that the signatures of two FCIs are also reproduced in device D1 (Fig. S5).

We pin down the $FCI_1$ in device D3 with a twist angle of 1.49°. Fig. 4A and 4B show color mappings of $R_{xx}(\nu,\mathrm{B})$ and $R_{xy}(\nu,\mathrm{B})$ at 500mK. White dashed lines are guidelines for the Chern insulator with $C$ = 3, and the pink dashed line is the one for the $FCI_1$ with $C$ = -6/5 at $\nu$ = ~12/5. The Chern number is obtained by tracing the evolution of Hall resistance peaks according to Streda's formula analysis (red dashed line in Fig. 4A). The same slope can also be identified in $R_{xx}$ mapping (red dashed line in Fig. 4B). We double-check the fractional Chern number from the quantized Hall resistance of $R_{xy} = -5h/(6e^2)$ at $B$ = 100mT and $T$ = 100mK in Fig. 4C, benchmarked by the quantization of the Chern insulator with $C$ = 3. Note that we could also identify a small $R_{xx}$ dip near $\nu$ = 12/5 together with a weak activation behavior (Fig. S7), suggesting a gap state for the fractionally quantized $R_{xy}$ at $\nu$ = 12/5; however, the $R_{xx}$ is quite large, and it might suffer from imperfect contact or domain boundary issues. Moreover, we observe fractional quantum anomalous Hall effects with a fractional Chern number of $C$ = -6/5 and a coercive magnetic field of 20mT (Fig. 4D). The anomalous Hall hysteresis window quickly thermal smears, and it is barely visible at T =820mK. As for the amplitude, compared to the integer Chern insulator at $\nu$ = 3 that could be quantized up to 3K, fQAH at $\nu$ = 12/5 deviates from fractional quantization at $T$ > 300mK (Fig. 4E).

Considering cascade transition at integer fillings and the trivial topology at $\nu$ = 2, we could define a true filling factor $\tilde{\nu}$ for the FCI observed in the third subband via $\tilde{\nu} = \nu - 2$. In this terminology, the $FCI_1$ with fQAH is observed with $C$ = -6/5 at $\tilde{\nu}$ = 2/5, denoted as $C_{2/5}$ = -6/5, and the parent integer Chern insulator is observed with $C$ = -3 at $\tilde{\nu}$ = 1, denoted $C_1$ = -3. And eventually, we obtain a linear relationship between Chern number and filling fraction, i.e., $C_{\tilde{\nu}} = \tilde{\nu}*C_1$. Note that the $FCI_2$ with $C$=-3/2 at $\nu$ = 5/2 from Streda's formula in device D2 follows the same linear rule, $C_{1/2}=1/2*C_1$.

**Discussions and Conclusion**

The above observations provide compelling experimental evidence of the fractional Chern insulators in the high Chern number flat band. Our observation of $FCI_1$ with $C_{2/5}$ = -6/5 is novel and distinct from all other FCIs based on realistic flatband systems with $C$ = 1 (*10, 13, 23, 24, 26*), beyond the conventional fractional quantum Hall effects in the Landau level paradigm. The $FCI_1$ with $C_{2/5}$ = -6/5 is unconventional and unexpected even theoretically. Recently, a theoretical work (*59*) extensively investigated isolated flatbands with high Chern numbers in tRTBG; however, there is still no realistic prediction of FCI. In addition, according to a lattice toy model for high Chern number flatbands (*31, 60*), the ground FCI states in flatbands with $C$ = 3 are expected to emerge at fractional fillings 1/7 for fermionic states or 1/4 for bosonic states; our observation of FCI at 2/5 is beyond the prediction. Intriguingly, FCI is also predicted in a recent theory for an isolated flatband with zero Chern number (*61*). In short, the underlying FCI mechanism is still challenging both for theory and experiments. More works are needed to address the issue, which, however, is beyond the scope of the current study.

To summarize, we have demonstrated that the twisted rhombohedral trilayer-bilayer graphene moiré system is a new ideal platform for developing novel zero-field topological phases with high Chern numbers at integer and fractional fillings, as shown schematically in Fig. S11. We observe QAH with a high Chern number $C = 3$ at both $\nu = 1$ and $\nu = 3$ in three separate devices. In addition, QAH with $C = 2$ is observed at even-denominator fractional filling $\nu = 3/2$, and more intriguingly, those with high Chern numbers $C = 3$ are observed developing continuously from $\nu = 1$ to 3/2. Most strikingly and importantly, for the first time, fQAH with $C_{2/5} = -6/5$ is observed at $\nu = 12/5$, corresponding to an FCI at the fractional filling 2/5 of the parent flat band with $C = -3$. Signature of a second FCI with $C_{1/2} = -3/2$ at even-denominator filling $\nu = 5/2$ is revealed from Streda's formula analysis, whose quantization is to be confirmed in future studies. These observations demonstrate that tRTBG offers new opportunities for engineering novel integer and fractional topological phases with high Chern numbers. The unconventional exotic phases in tRTBG can be naturally extended to other twisted graphene moiré superlattices based on rhombohedral multilayers. More experimental and theoretical efforts are needed to address the fractional phases in the high Chern number flatbands that could be of non-Abelian statistics (*62, 63*) for future topological quantum computing.

**References and Notes**

1. F. D. Haldane, Model for a quantum Hall effect without Landau levels: Condensed-matter realization of the "parity anomaly". *Physical Review Letters* **61**, 2015-2018 (1988).
2. C. Z. Chang, C. X. Liu, A. H. MacDonald, Colloquium: Quantum anomalous Hall effect. *Reviews of Modern Physics* **95**, 011002 (2023).
3. C.-Z. Chang *et al.*, Experimental Observation of the Quantum Anomalous Hall Effect in a Magnetic Topological Insulator. *Science* **340**, 167-170 (2013).
4. Y. Deng *et al.*, Quantum anomalous Hall effect in intrinsic magnetic topological insulator $MnBi_2Te_4$. *Science* **367**, 895-900 (2020).
5. A. L. Sharpe *et al.*, Emergent ferromagnetism near three-quarters filling in twisted bilayer graphene. *Science* **365**, 605-608 (2019).
6. M. Serlin *et al.*, Intrinsic quantized anomalous Hall effect in a moire heterostructure. *Science* **367**, 900-903 (2020).
7. G. Chen *et al.*, Tunable correlated Chern insulator and ferromagnetism in a moire superlattice. *Nature* **579**, 56-61 (2020).
8. H. Polshyn *et al.*, Electrical switching of magnetic order in an orbital Chern insulator. *Nature* **588**, 66-70 (2020).
9. T. Li *et al.*, Quantum anomalous Hall effect from intertwined moire bands. *Nature* **600**, 641-646 (2021).
10. Z. Lu *et al.*, Fractional quantum anomalous Hall effect in multilayer graphene. *Nature* **626**, 759-764 (2024).
11. T. Han *et al.*, Large quantum anomalous Hall effect in spin-orbit proximitized rhombohedral graphene. *Science* **384**, 647-651 (2024).
12. Y. Sha *et al.*, Observation of a Chern insulator in crystalline ABCA-tetralayer graphene with spin-orbit coupling. *Science* **384**, 414-419 (2024).
13. J. Xie *et al.*, Tunable fractional Chern insulators in rhombohedral graphene superlattices.

*Nature Materials*, (2025).
14. Y. Choi *et al.*, Superconductivity and quantized anomalous Hall effect in rhombohedral graphene. *Nature* **639**, 342-347 (2025).
15. R. Su *et al.*, Moiré-driven topological electronic crystals in twisted graphene. *Nature* **637**, 1084-1089 (2025).
16. T. Neupert, L. Santos, C. Chamon, C. Mudry, Fractional quantum Hall states at zero magnetic field. *Physical Review Letters* **106**, 236804 (2011).
17. K. Sun, Z. Gu, H. Katsura, S. Das Sarma, Nearly Flatbands with Nontrivial Topology. *Physical Review Letters* **106**, 236803 (2011).
18. E. Tang, J.-W. Mei, X.-G. Wen, High-Temperature Fractional Quantum Hall States. *Physical Review Letters* **106**, 236802 (2011).
19. N. Regnault, B. A. Bernevig, Fractional Chern Insulator. *Physical Review X* **1**, 021014 (2011).
20. D. N. Sheng, Z. C. Gu, K. Sun, L. Sheng, Fractional quantum Hall effect in the absence of Landau levels. *Nature Communications* **2**, 389 (2011).
21. D. C. Tsui, H. L. Stormer, A. C. Gossard, Two-Dimensional Magnetotransport in the Extreme Quantum Limit. *Physical Review Letters* **48**, 1559-1562 (1982).
22. R. Willett *et al.*, Observation of an even-denominator quantum number in the fractional quantum Hall effect. *Physical Review Letters* **59**, 1776-1779 (1987).
23. J. Cai *et al.*, Signatures of fractional quantum anomalous Hall states in twisted $MoTe_2$. *Nature* **622**, 63-68 (2023).
24. H. Park *et al.*, Observation of fractionally quantized anomalous Hall effect. *Nature* **622**, 74-79 (2023).
25. Y. Zeng *et al.*, Thermodynamic evidence of fractional Chern insulator in moire $MoTe_2$. *Nature* **622**, 69-73 (2023).
26. F. Xu *et al.*, Observation of Integer and Fractional Quantum Anomalous Hall Effects in Twisted Bilayer $MoTe_2$. *Physical Review X* **13**, 031037 (2023).
27. H. Bartolomei *et al.*, Fractional statistics in anyon collisions. *Science* **368**, 173-177 (2020).
28. J. Nakamura, S. Liang, G. C. Gardner, M. J. Manfra, Direct observation of anyonic braiding statistics. *Nature Physics* **16**, 931-936 (2020).
29. X. G. Wen, Non-Abelian statistics in the fractional quantum Hall states. *Physical Review Letters* **66**, 802-805 (1991).
30. C. Nayak, S. H. Simon, A. Stern, M. Freedman, S. Das Sarma, Non-Abelian anyons and topological quantum computation. *Reviews of Modern Physics* **80**, 1083-1159 (2008).
31. Z. Liu, E. J. Bergholtz, H. Fan, A. M. Läuchli, Fractional Chern Insulators in Topological Flat Bands with Higher Chern Number. *Physical Review Letters* **109**, 186805 (2012).
32. A. Sterdyniak, C. Repellin, B. A. Bernevig, N. Regnault, Series of Abelian and non-Abelian states in $C > 1$ fractional Chern insulators. *Physical Review B* **87**, 205137 (2013).
33. F. Zhang, B. Sahu, H. Min, A. H. MacDonald, Band structure ofABC-stacked graphene trilayers. *Physical Review B* **82**, 035409 (2010).
34. N. Bultinck, S. Chatterjee, M. P. Zaletel, Mechanism for Anomalous Hall Ferromagnetism in Twisted Bilayer Graphene. *Physical Review Letters* **124**, 166601 (2020).
35. F. Wu, S. Das Sarma, Collective Excitations of Quantum Anomalous Hall Ferromagnets in Twisted Bilayer Graphene. *Physical Review Letters* **124**, 046403 (2020).
36. H.-K. Tang *et al.*, The role of electron-electron interactions in two-dimensional Dirac fermions. *Science* **361**, 570-574 (2018).
37. M. Koshino, E. McCann, Trigonal warping and Berry's phase N $\pi$ in ABC-stacked multilayer graphene. *Physical Review B* **80**, 165409 (2009).

38. A. Abouelkomsan, Z. Liu, E. J. Bergholtz, Particle-Hole Duality, Emergent Fermi Liquids, and Fractional Chern Insulators in Moiré Flatbands. *Physical Review Letters* **124**, 106803 (2020).
39. Z. Lu *et al.*, Extended quantum anomalous Hall states in graphene/hBN moire superlattices. *Nature* **637**, 1090-1095 (2025).
40. Y.-H. Zhang, D. Mao, Y. Cao, P. Jarillo-Herrero, T. Senthil, Nearly flat Chern bands in moiré superlattices. *Physical Review B* **99**, 075127 (2019).
41. J. Liu, Z. Ma, J. Gao, X. Dai, Quantum Valley Hall Effect, Orbital Magnetism, and Anomalous Hall Effect in Twisted Multilayer Graphene Systems. *Physical Review X* **9**, 031021 (2019).
42. P. J. Ledwith, A. Vishwanath, E. Khalaf, Family of Ideal Chern Flatbands with Arbitrary Chern Number in Chiral Twisted Graphene Multilayers. *Physical Review Letters* **128**, 176404 (2022).
43. J. Wang, Z. Liu, Hierarchy of Ideal Flatbands in Chiral Twisted Multilayer Graphene Models. *Physical Review Letters* **128**, 176403 (2022).
44. C. L. Tschirhart *et al.*, Imaging orbital ferromagnetism in a moire Chern insulator. *Science* **372**, 1323-1327 (2021).
45. J. Zhu, J. J. Su, A. H. MacDonald, Voltage-Controlled Magnetic Reversal in Orbital Chern Insulators. *Physical Review Letters* **125**, 227702 (2020).
46. H. Polshyn *et al.*, Topological charge density waves at half-integer filling of a moiré superlattice. *Nature Physics* **18**, 42-47 (2021).
47. X.-Y. Song, C.-M. Jian, L. Fu, C. Xu, Intertwined fractional quantum anomalous Hall states and charge density waves. *Physical Review B* **109**, 115116 (2024).
48. D. N. Sheng, A. P. Reddy, A. Abouelkomsan, E. J. Bergholtz, L. Fu, Quantum Anomalous Hall Crystal at Fractional Filling of Moiré Superlattices. *Physical Review Letters* **133**, 066601 (2024).
49. W. Yang *et al.*, Fractional Quantum Anomalous Hall Effect in a Singular Flat Band. *Physical Review Letters* **134**, 196501 (2025).
50. R. Perea-Causin, H. Liu, E. J. Bergholtz, Quantum anomalous Hall crystals in moiré bands with higher Chern number. arXiv:2412.02745 (2024).
51. B. Zhou, Y.-H. J. a. e.-p. Zhang, New classes of quantum anomalous Hall crystals in multilayer graphene. arXiv:2411.04174 (2024).
52. S. Kivelson, C. Kallin, D. P. Arovas, J. R. Schrieffer, Cooperative ring exchange theory of the fractional quantized Hall effect. *Physical Review Letters* **56**, 873-876 (1986).
53. B. I. Halperin, Z. Tešanović, F. Axel, Compatibility of Crystalline Order and the Quantized Hall Effect. *Physical Review Letters* **57**, 922-922 (1986).
54. Z. Tešanović, F. Axel, B. I. Halperin, ‘‘Hall crystal’’ versus Wigner crystal. *Physical Review B* **39**, 8525-8551 (1989).
55. T. Tan, T. Devakul, Parent Berry Curvature and the Ideal Anomalous Hall Crystal. *Physical Review X* **14**, 041040 (2024).
56. J. Dong *et al.*, Anomalous Hall Crystals in Rhombohedral Multilayer Graphene. I. Interaction-Driven Chern Bands and Fractional Quantum Hall States at Zero Magnetic Field. *Physical Review Letters* **133**, 206503 (2024).
57. F. Xu *et al.*, Signatures of unconventional superconductivity near reentrant and fractional quantum anomalous Hall insulators. arXiv:2504.06972 (2025).
58. L. Liu *et al.*, Observation of First-Order Quantum Phase Transitions and Ferromagnetism in Twisted Double Bilayer Graphene. *Physical Review X* **13**, 031015 (2023).
59. Vo Tien Phong, C. Lewandowski, Coulomb Interaction-Stabilized Isolated Narrow Bands

with Chern Numbers C > 1 in Twisted Rhombohedral Trilayer-Bilayer Graphene. arXiv:2505.07981 (2025).
60. G. Möller, N. R. Cooper, Fractional Chern Insulators in Harper-Hofstadter Bands with Higher Chern Number. *Physical Review Letters* **115**, 126401 (2015).
61. Z. Lin, H. Lu, W. Yang, D. Zhai, W. Yao, Fractional Chern insulator states in an isolated flat band of zero Chern number. arXiv:2505.09009 (2025).
62. A. P. Reddy, N. Paul, A. Abouelkomsan, L. Fu, Non-Abelian Fractionalization in Topological Minibands. *Physical Review Letters* **133**, 166503 (2024).
63. H. Liu, Z. Liu, E. J. J. a. e.-p. Bergholtz, Non-Abelian Fractional Chern Insulators and Competing States in Flat Moiré Bands. arXiv:2405.08887 (2024).
64. Z. Feng *et al.*, Rapid infrared imaging of rhombohedral graphene. *Physical Review Applied* **23**, 034012 (2025).
65. L. Wang *et al.*, One-Dimensional Electrical Contact to a Two-Dimensional Material. *Science* **342**, 614-617 (2013).
66. E. Redekop *et al.*, Direct magnetic imaging of fractional Chern insulators in twisted MoTe2. *Nature* **635**, 584-589 (2024).

**Acknowledgments:** We thank Xiaobo Lu and Zihao Huo for their assistance in the sample fabrication of rhombohedral graphene. We also thank Wang Yao, Xi Dai, Jianpeng Liu, Nicolas Regnault, Xiao Li, Xincheng Xie, Jie Shan, Kinfai Mak, and Gwendal Feve for useful discussions.

**Funding:** We acknowledge support from the National Key Research and Development Program (Grant Nos. 2020YFA0309600, 2024YFA1410400, 2021YFA1202900), the Beijing Natural Science Foundation (Grant No. JQ25003), the Natural Science Foundation of China (NSFC, Grant No. 62488201), and the Guangdong Major Project of Basic and Applied Basic Research (Grant No. 2021B0301030002). K.W. and T.T. acknowledge support from the JSPS KAKENHI (Grant Numbers 20H00354, 21H05233 and 23H02052) and World Premier International Research Center Initiative (WPI), MEXT, Japan.

**Author contributions:** W.Y. and G.Z. supervised the project; W.Y. and J.D. designed the experiments; J.D. fabricated the devices and performed the magneto-transport measurement with the help of J.Z., Z.P., Y.H., Z.J.; L.L. performed the calculations. K.W. and T.T. provided hexagonal boron nitride crystals. W.Y., J.D., L.L., and G.Z. analyzed the data. W.Y. wrote the paper with the input of J.D., L.L., and G.Z.; all authors commented on the manuscript.

**Competing interests:** The authors declare no competing interests.

**Data and materials availability:** The data that support the findings of this study are available from the corresponding authors upon reasonable request.

**Supplementary Materials**

Materials and Methods

Supplementary Text

Figs. S1 to S11

References (*64–66*)

# Table

| Dev. | $\theta$ | Top Gate | Chern number ($C$) extracted from Streda's formula analysis | | | | | | |
|---|---|---|---|---|---|---|---|---|---|
| | | | $\nu = 1/2$ | $\nu = 1$ | $1 < \nu < 1.5$ | $\nu = 3/2$ | $\nu = 12/5$ | $\nu = 5/2$ | $\nu = 3$ |
| D1 | 1.35±0.01° | Graphite | \ | 3 (**QAH**) | 3 (**QAH**) | 2 (**QAH**) | -6/5 | \ | -3 (**QAH**) |
| D2 | 1.5±0.02° | Graphite | Topological phases at $\nu<2$ are not accessible due to a technical issue of gate leakage. | | | | -6/5 | -3/2 | -3 (**QAH**) |
| D3 | 1.49±0.01° | Graphite | \ | 3 (**QAH**) | 3 (**QAH**) | 2 (**QAH**) | -6/5 (**fQAH**) | \ | -3 (**QAH**) |
| D4 | 1.42° | Metal | 1 | 3 | \ | 2 | \ | \ | 3 |
| D5 | 1.3° | Metal | 1 | 3 | \ | 2 | \ | \ | 3 |
| D6 | 1.32° | Metal | \ | 3 | \ | \ | \ | \ | 3 |
| D7 | 1.28° | Metal | \ | 3 | \ | \ | \ | \ | \ |

**Table. 1. A summary of experimental parameters and results observed in different tRTBG devices.**

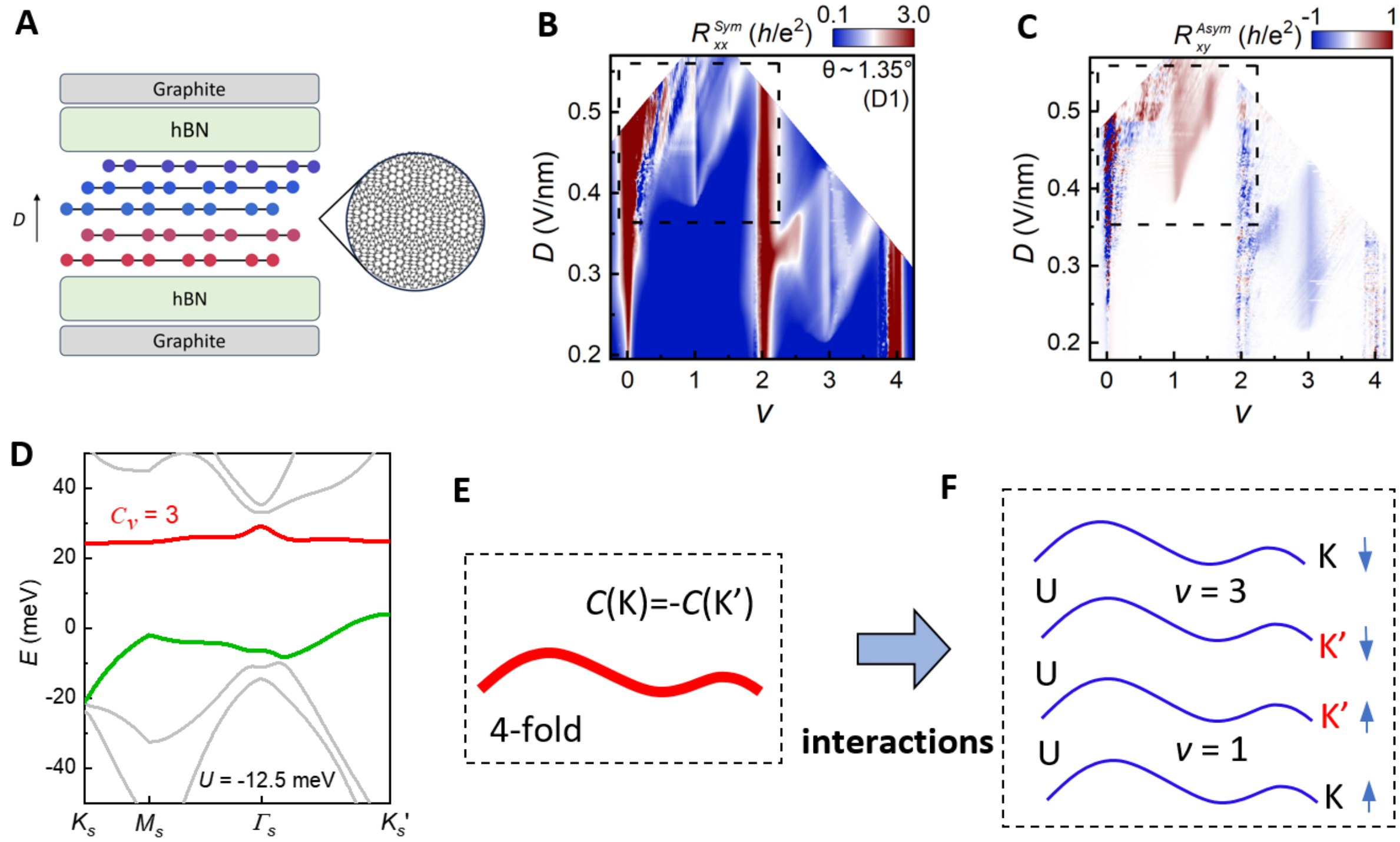

**Fig. 1. High Chern number flatbands in tRTBG.** (**A**) Schematic of the tRTBG sample structure. The illusion shows a moiré superlattice between rhombohedral trilayer graphene and Bernal bilayer graphene. The positive D field direction is defined as pointing from the bilayer towards the rhombohedral trilayer. (**B** and **C**) Color mappings of $R_{xx}$ (B) and $R_{xy}$ (C) as a function of displacement field $D$ and filling factor $\nu$ at $T = 1$ K in device D1 with $\theta = 1.35°$. (**D**) Calculated band structures of tRTBG with $\theta = \sim 1.3°$ at interlayer potential $U = -12.5$ meV, showing an isolated flat conduction band with valley Chern number of $C_v = 3$ (red curve). (**E** and **F**) Schematics of the moiré conduction flatband with four-fold degeneracy (E). Coulomb interactions lift the flatband degeneracy, resulting in spin and valley polarized subbands and correlated insulators at integer fillings (F). The isospin polarizations are chosen according to the experimental data in (C).

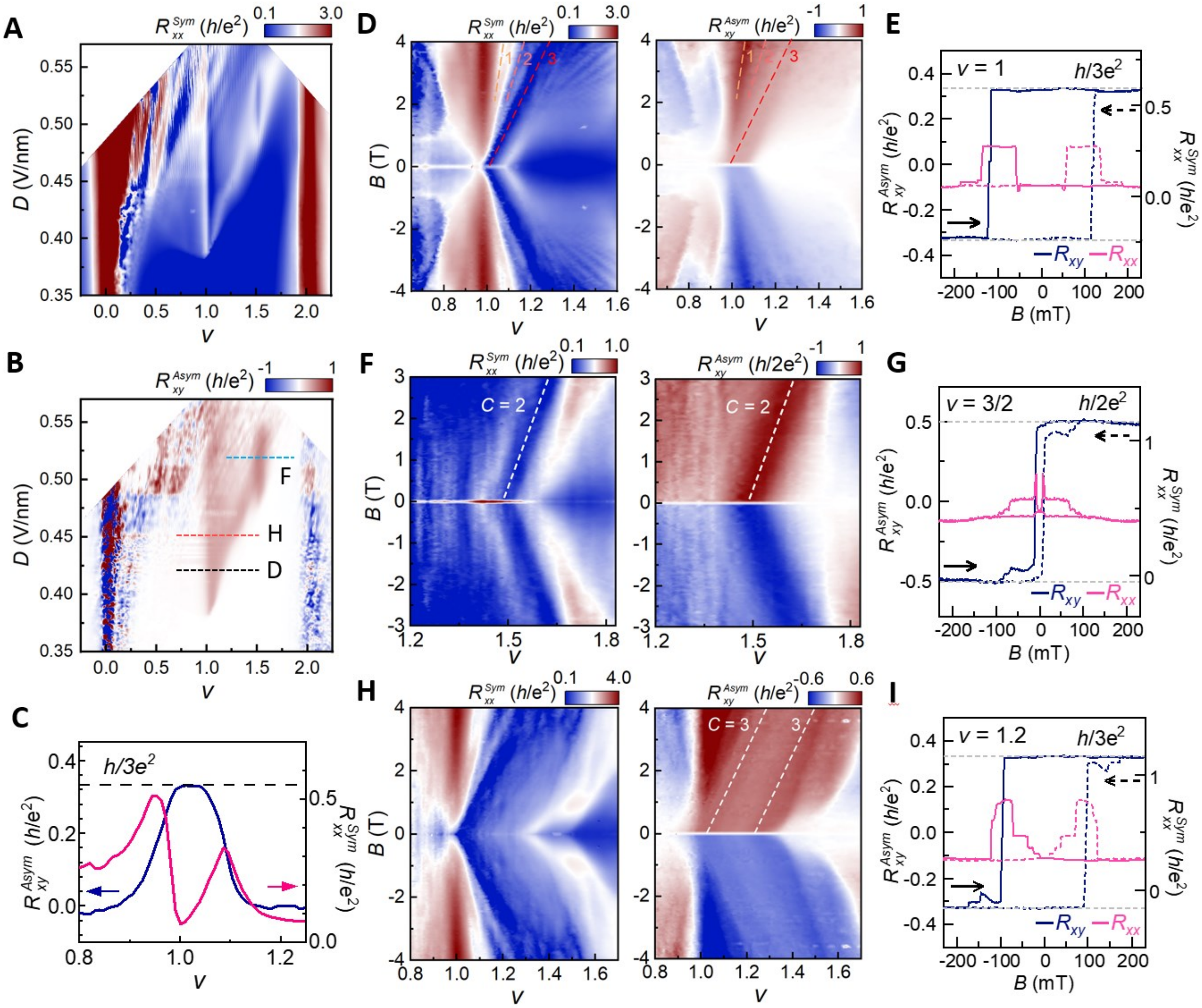

**Fig. 2. High Chern number integer Quantum anomalous Hall effects at integer and fractional fillings near $\nu = 1$.** (**A** and **B**) Phase diagram of symmetrized $R_{xx}$ (A) and anti-symmetrized $R_{xy}$ (B) at $B = \pm 0.1$ T. (**C**) Plots to show quantized Hall plateau with $R_{xy} = h/(3e^2)$ and a $R_{xx}$ minimum across $\nu = 1$. (**D**) Landau fan diagram along the black dashed line in (B). The red dashed line originating from $B = 0$ indicates a Chern insulator with $C = 3$ determined by Streda's formula. (**E**) Plots to show Quantum anomalous Hall effects (QAH) with $C = 3$ at $\nu = 1$ and $D = 0.42$ V/nm. (**F**) Landau fan diagram along the blue dashed line in (B) across $\nu = 3/2$. The dashed line indicates a Chern insulator with $C = 2$ determined by Streda's formula. (**G**) Plots to show QAH with $C = 2$ at $\nu = 3/2$ and $D = 0.515$ V/nm. (**H**) Landau fan diagram along the red dashed line in (B). The dashed lines indicate Chern insulators with $C = 3$ determined by Streda's formula. (**I**) Plots to show QAH with $C = 3$ at $\nu = 1.2$ and $D = 0.443$ V/nm. Data are obtained from device D1 at $T = 100$mK.

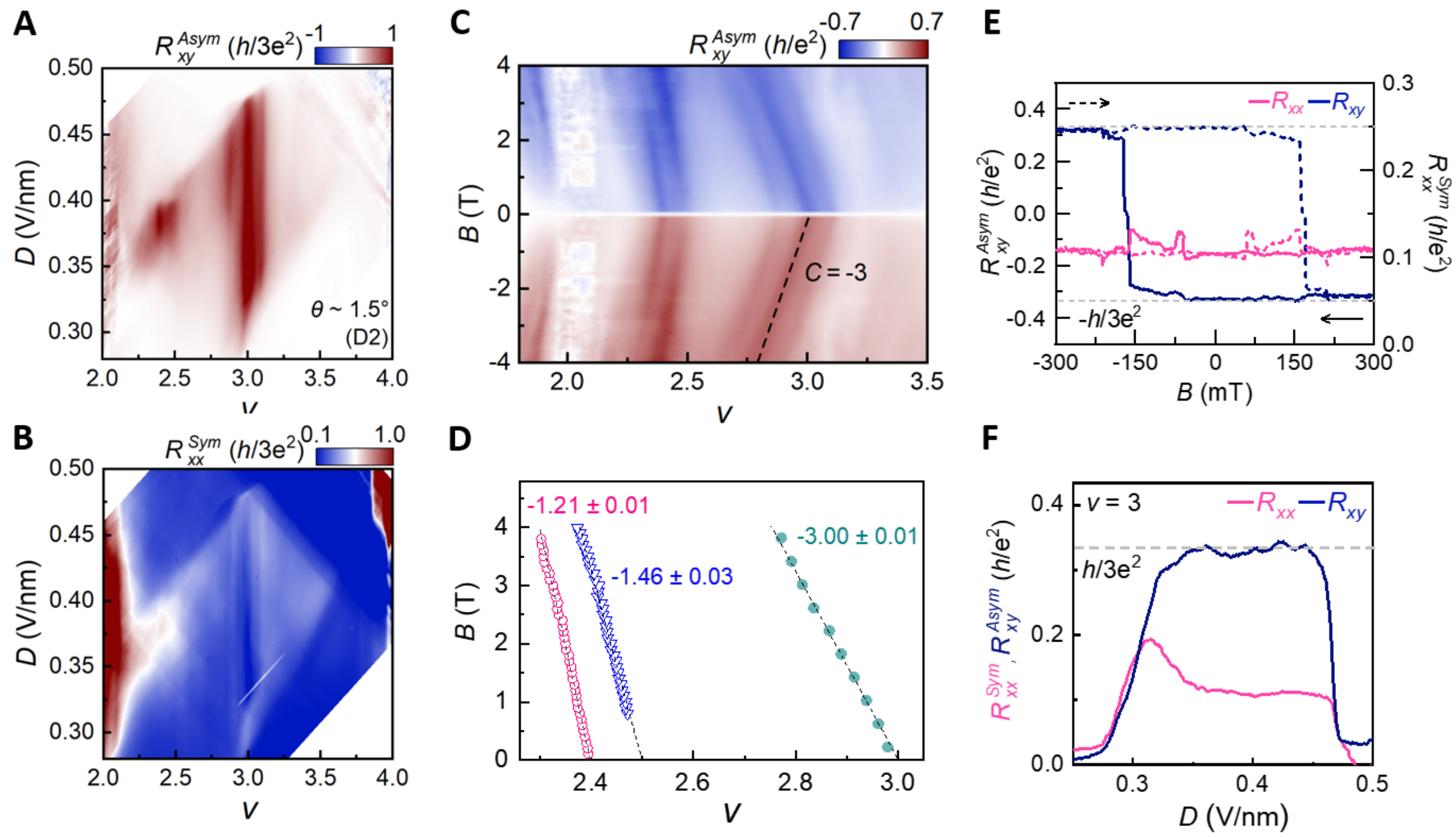

**Fig. 3. High Chern number QAH with $C$ = -3 at $\nu$ = 3.** (**A** and **B**) Phase diagrams of anti-symmetrized $R_{xy}$ (A) and symmetrized $R_{xx}$ (B) as a function of displacement field $D$ and filling factor $\nu$ in device D2 with $\theta$ = 1.5°at $B$ = -0.1 T. (**C**) Landau fan diagram of $R_{xy}$ at $D$ = 0.384 V/nm. The black dashed line indicates the Chern insulator with $C$ = -3. (**D**) Streda's formula analysis suggests a Chern insulator with $C$ = -3 at $\nu$ = 3. It also suggests two fractional Chern insulators, one with $C$ = -1.21 ± 0.01 at $\nu$ = 2.4 and the other with $C$ = -1.46 ± 0.03 at $\nu$ = 2.5. (**E**) Plots to show QAH with $C$ = -3 at $\nu$ = 3 and $D$ = 0.408 V/nm. (**F**) $R_{xy}$ and $R_{xx}$ as a function of $D$ at $\nu$ = 3 and $B$ = -0.1T. The data are measured at $T$ = 65mK.

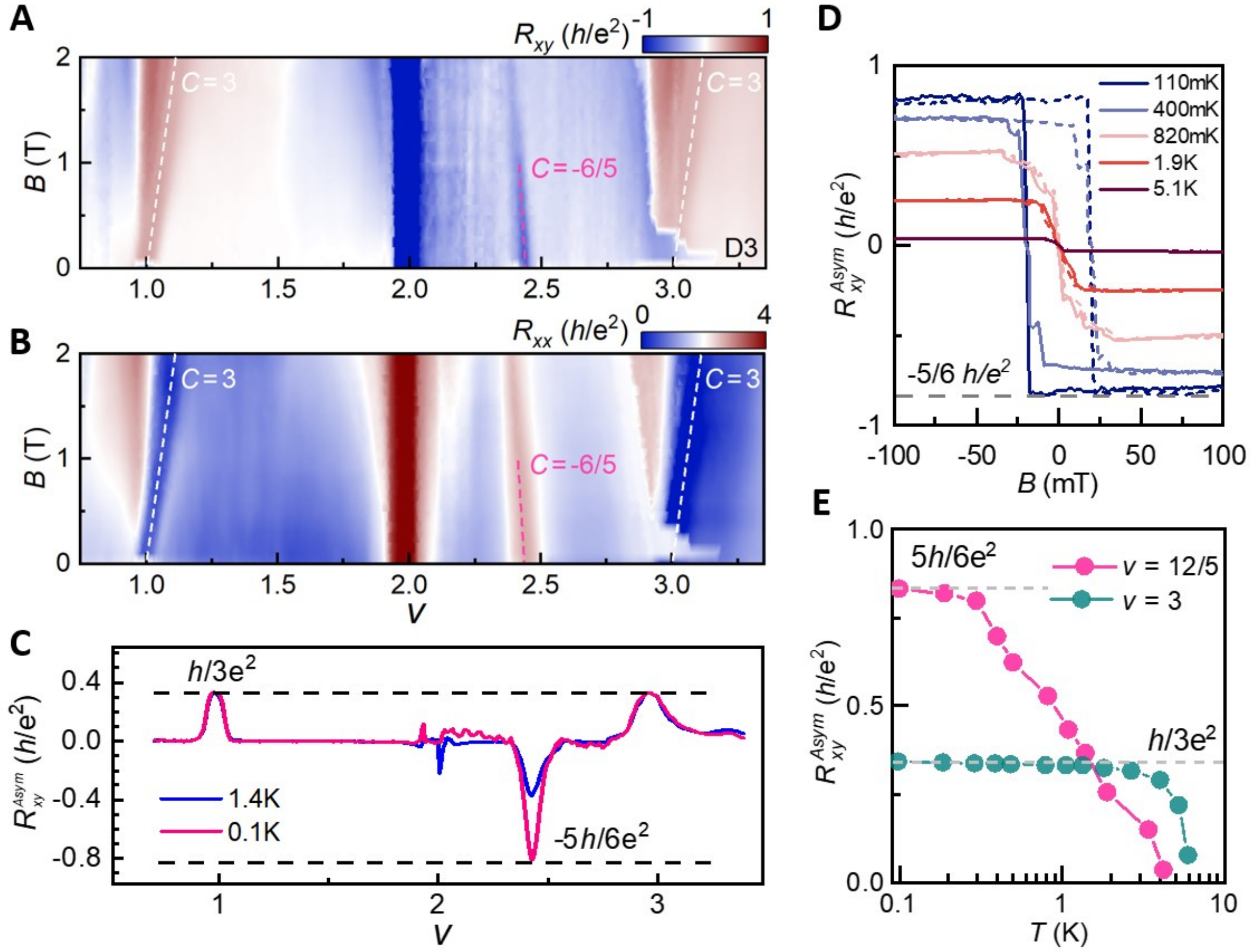

**Fig. 4. Fractional Chern insulator with $C$ = -6/5 at $v$ = 12/5.** (**A** and **B**) Landau fan diagrams to show integer Chern insulators and fractional Chern insulators along the black dashed line in Fig. S**7**A at $T$ = 500mK in device D3 with $\theta$ = 1.49°. The white dashed lines indicate the Chern insulator with $C$ = 3 at $v$ = 1 and $v$ = 3, and the pink ones indicate the fractional Chern insulator (FCI) with $C$ = -6/5 at $v$ = ~2.4. (**C**) Plots of anti-symmetrized $R_{xy}(v)$ at $B$ = 0.1T. The FCI reaches the quantization of $R_{xy}$ = -5$h$/6e$^2$ at $T$ = 100mK (pink), while the Chern insulators at $v$ = 1 and 3 keep the quantization of $R_{xy}$ = $h$/3e$^2$ up to $T$ = 1.4K (blue). (**D**) Fractional quantum anomalous Hall effects of the FCI at different temperatures. (**E**) Temperature dependence of $R_{xy}$ plateau for the Chern insulator at $v$ = 3 and FCI at $v$ = 12/5.

# Supplementary Information

## Materials and Methods

Device fabrication

We fabricate tRTBG samples by the cut-and-stacking method. We first exfoliate graphene multilayers on 300nm $SiO_2$/Si substrate with scotch tape and use the infrared imaging technique (*64*) and Raman spectroscopy to identify the rhombohedral trilayer graphene. We elaborately focus on the graphene flakes with bilayer and rhombohedral trilayer graphene on the same flake and cut them into two pieces, one for bilayer graphene and the other for trilayer graphene. Then, we use the dry pick-up technique to stack the top hBN, bilayer graphene, trilayer graphene, bottom hBN, and few-layer graphite (as bottom gate electrode) flakes sequentially by using polypropylene carbonate (PPC) film. The stacked samples are fabricated into dual gate Hall bar structures following the standard e-beam lithography, reactive-ion etching, and e-beam evaporation of Cr - Au (edge contact(*65*) and top gate electrode) processes.

Electrical transport measurements

The devices are measured in a Janis helium-4 cryostat (base temperature ~ 1.7 K) and a Leiden dilution refrigerator (base temperature ~ 20 mK) with a standard low-frequency lock-in technique. We measure the four terminal longitudinal resistance $R_{xx}$ and Hall resistance $R_{xy}$ using the lock-in amplifier SR830/LI5650 with a frequency of 10~35 Hz and excitation currents of 1-10 nA through a large series resistance 100 MΩ. The top and the bottom gates are supplied by Keithley 2612 and Keithley 2400 source meters.

The dual gate structures allow independent control of the carrier density ($n$) and the electric displacement field ($D$), i.e. $n = (C_{BG}V_{BG} + C_{TG}V_{TG})/e$ and $D = (C_{BG}V_{BG} - C_{TG}V_{TG})/2\varepsilon_0$. Here, $C_{BG}$ ($C_{TG}$) is the geometrical capacitance per area for the bottom (top) gate; $e$ is the electron charge; $\varepsilon_0$ is the vacuum permittivity. The filling factor is defined as $\nu = 4n/n_s$, where $n_s = 4/A$ is the carrier density at full fillings and $A$ is the area of the moiré supercell that depends on the twist angle θ. For small twist angles, $n_s \approx 8\theta^2/(\sqrt{3}a^2)$, where $a$ is the lattice constant of graphene.

Device fabrication

To eliminating the crosstalk between the longitudinal and transverse resistance, we perform the symmetric and anti-symmetric operating for the measured longitudinal resistance and Hall resistance respectively. That is $R_{xx}^{sym}(\pm B) = (R_{xx}(\pm B) + R_{xx}(\mp B))/2$ and $R_{xy}^{asym}(\pm B) = (R_{xy}(\pm B) - R_{xy}(\mp B))/2$.

## Supplementary Text

Band structure and Chern number calculations

The energy band calculations are performed with a continuum model. Considering the two stacking orders of trilayer graphene, i.e. ABA and ABC, we calculate the band structures for both AB-ABA and AB-ABC configurations. The Hamiltonian of trilayer graphene is written in the basis ($A_1$, $B_1$, $A_2$, $B_2$, $A_3$, $B_3$). Here $A_i$ and $B_i$ denote the A and B sublattice sites of the *i* layer. The parameters we used are ($\gamma_0$, $\gamma_1$, $\gamma_2$, $\gamma_3$, $\gamma_4$, $\gamma_5$) = (2610, 361, -20, 283, 141, 20) meV. Here $\gamma_0$ is the nearest intralayer hopping term; $\gamma_1$, $\gamma_3$ and $\gamma_4$ are the interlayer hopping terms between two adjacent layers; $\gamma_2$ and $\gamma_5$ (only in ABA-stacking) are the interlayer hopping terms between two outmost layers. We use the same parameters for AB stacking bilayer graphene. The coupling terms between two twisted layers are set as ($u_{AA}$, $u_{AB}$) = (80, 100) meV considering the corrugation effect. The potential energy $U$ is the energy difference between two adjacent layers. $U > 0$ corresponds to the direction of the electric field pointing from bilayer to trilayer.

We set the twisted angle at ~1.3° and calculate $U$-dependent energy bands for both AB-ABA and AB-ABC stacking. In both cases, there is an optimal isolated flat band condition at finite negative $U$ and a larger bandwidth at positive $U$. The AB-ABC configuration endows a better condition with an isolated flat band compared with the AB-ABA configuration. In addition, we calculate the valley Chern number for both configurations. The results are shown in Fig. S3, where the flat band in AB-ABC is found to be topologically nontrivial with a high valley Chern number of $C_v = 3$. For comparison, the flat band in AB-ABA is also topological, however, with a smaller valley Chern number of $C_v = 1$, as confirmed experimentally in ref. 15(*15*). The observation of $C_v = 3$ is in line with the experimental results and thus confirms the rhombohedral stacking order in our devices.

Degradation of sample quality

In many devices, the measured anomalous Hall resistances at quarter fillings deviate from quantization at $\rho_{xy} = h/(Ce^2)$. For instance, in our tRTBG device D4 with $\theta = 1.42^{\circ}$, Chern insulators with $C = 3$ and anomalous Hall effect are observed, yet $R_{xy}$ is saturated at ~ 2 kΩ. The observations hint at a degradation of device quality, possibly due to the partial relaxation of rhombohedral stacking to Bernal stacking or the increased strain and other imperfections that are detrimental to the size of the ferromagnetic domain(*66*). The reason can be twofold. One is the degradation of device quality during cooling down, since the strain and relaxation of the rhombohedral stacking may accumulate after thermal cycling. Indeed, we observe a degraded anomalous Hall effect when the device is remeasured in a second cooling down, as shown in the comparison between the two cooling down for device D2 in Fig. S8. The other, and also very crucial reason, might lie in the metal top gate. Since hBN is a good thermal conductor, the evaporation of metal directly on the surface of the top hBN layer could transfer the high energy of the evaporated metal particles to the tRTBG sample. Because of the metastable nature of the rhombohedral stacking, the metal top gate leads to a partial relaxation of rhombohedral stacking to Bernal stacking and the creation of multiple domains with reduced size and increased domain boundary; and eventually, it leads to a failure of quantization, as observed in many devices such as D4 and D5 with metal top gates.

Correlated Chern insulators in device D5

The tRBTG device D5, with a twist angle $\theta = 1.3^{\circ}$, shares the same structure as that in D4, with gold thin film as the top gate and a thin graphite layer as the bottom gate. The main experimental data are shown in Fig. S10. Here, we summarize the main observations in the following:

1) By performing the Streda formula analysis as indicated by the dashed lines (Fig. S10A), we obtain a high Chern number of $C = 3$ at integer fillings $\nu = 1$ and 3 (Fig. S10B-D), evident from the measured Hall resistance plateau that onsets at zero magnetic fields with a corresponding $R_{xx}$ minimum. These observations agree with the Chern insulators with $C = 3$ at $\nu = 1$ and 3 in device D1 and the band topology calculations.

2) We observe a Chern insulator with $C = 1$ at even-denominator fractional filling $\nu = 1/2$ in a limited D range from 0.55 to 0.6V/nm, verified according to the Streda relationship in the Landau fan diagram taken along the black dashed line at $D = 0.575$ V/nm (Fig. S10B).

3) We observe a series of topological phases with negative Chern number $C = -2$ developed at $\nu < 1$ and high magnetic fields $B > 1.3$T (Fig. S10C), accompanied by first-order quantum phase transitions.

4) Similar quantum phase transitions are also observed with negative Chern number $C = -2$ developed at $\nu < 3$ and $B > 0.5$T (Fig. S10E-F). The coexistence of Chern insulators with opposite signs at high magnetic fields suggests a complicated phase diagram with a coexistence of quasi-hole and quasi-electron topological phases.

In addition, it features a prominent hysteresis of the Hall resistance as the carrier density is swept back and forth from $\nu = 2.7$ to $\nu = 3.2$ at $B < B_c = 0.5$T, while the hysteresis disappears at $B > 0.5$T. The hysteresis with opposite sign indicates two metastable valley polarizations for the zero-field Chern insulators, and the valley polarization can be further stabilized under high magnetic fields.

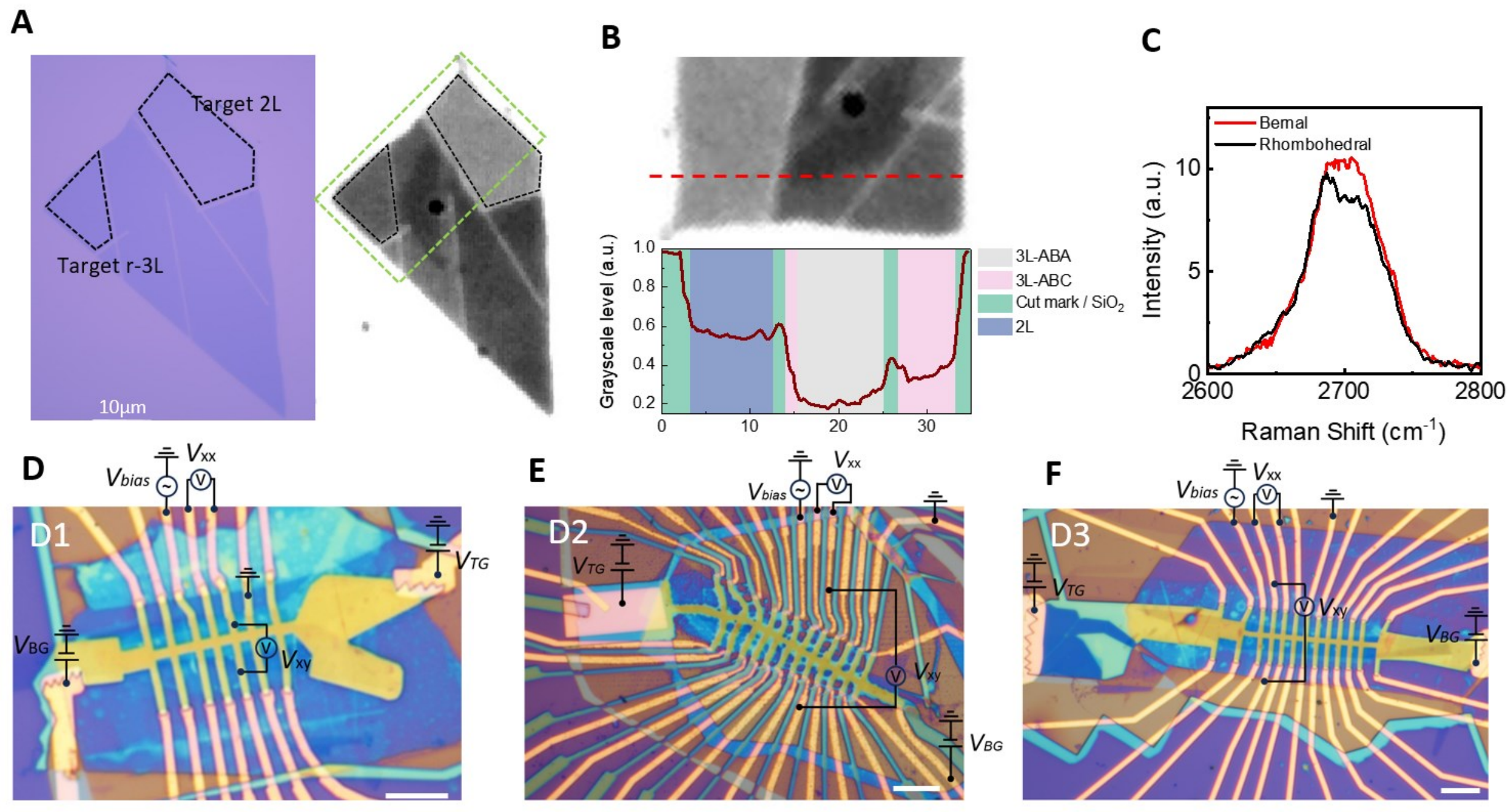

**Fig. S1.**

**Evidence of rhombohedral stacking and optical micrographs of tRTBG devices.** (**A**) Optical image (left panel) and infrared image (right panel) of exfoliated graphene. The target bilayer and trilayer graphene is indicated by black dashed box. (**B**) Zoom-in infrared image (upper panel), corresponding to green dashed box in (A)**.** Grayscale diagram (lower panel) extracted from line cut in infrared image (red dashed line). In the infrared images, the ABC region appears lighter in color compared to the ABA region, and it exhibits larger values than ABA in the grayscale diagram. (**C**) Raman spectroscopy measured at rhombohedral and bernal trilayer graphene sample in (A). (**D-F**) optical micrographs of device D1 (D), D2 (E), and D3 (F) used for transport measurements in the main text. The scale bar is 5 um.

twisted rhombohedral trilayer-bilayer graphene

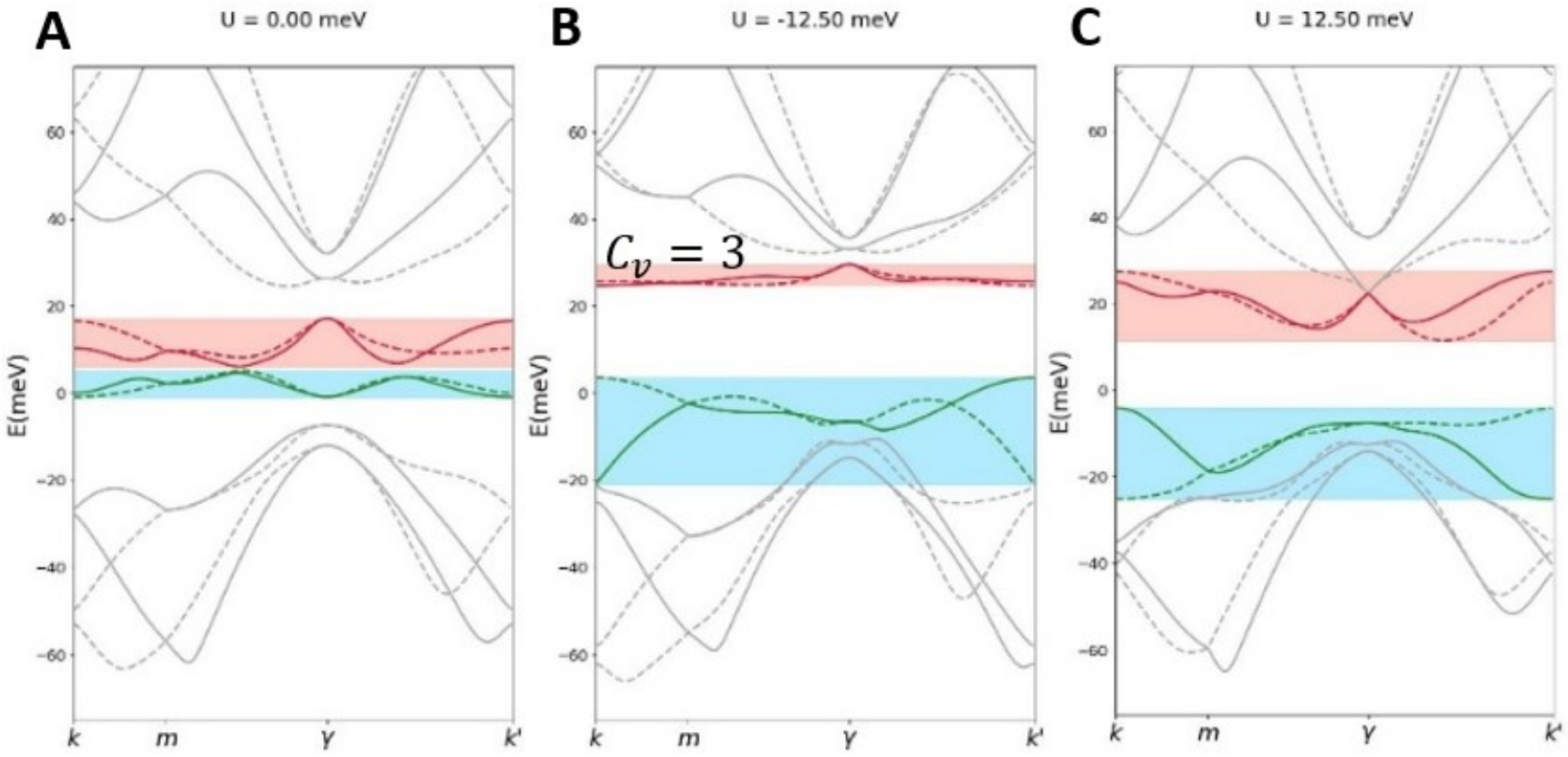

twisted Bernal trilayer-bilayer graphene

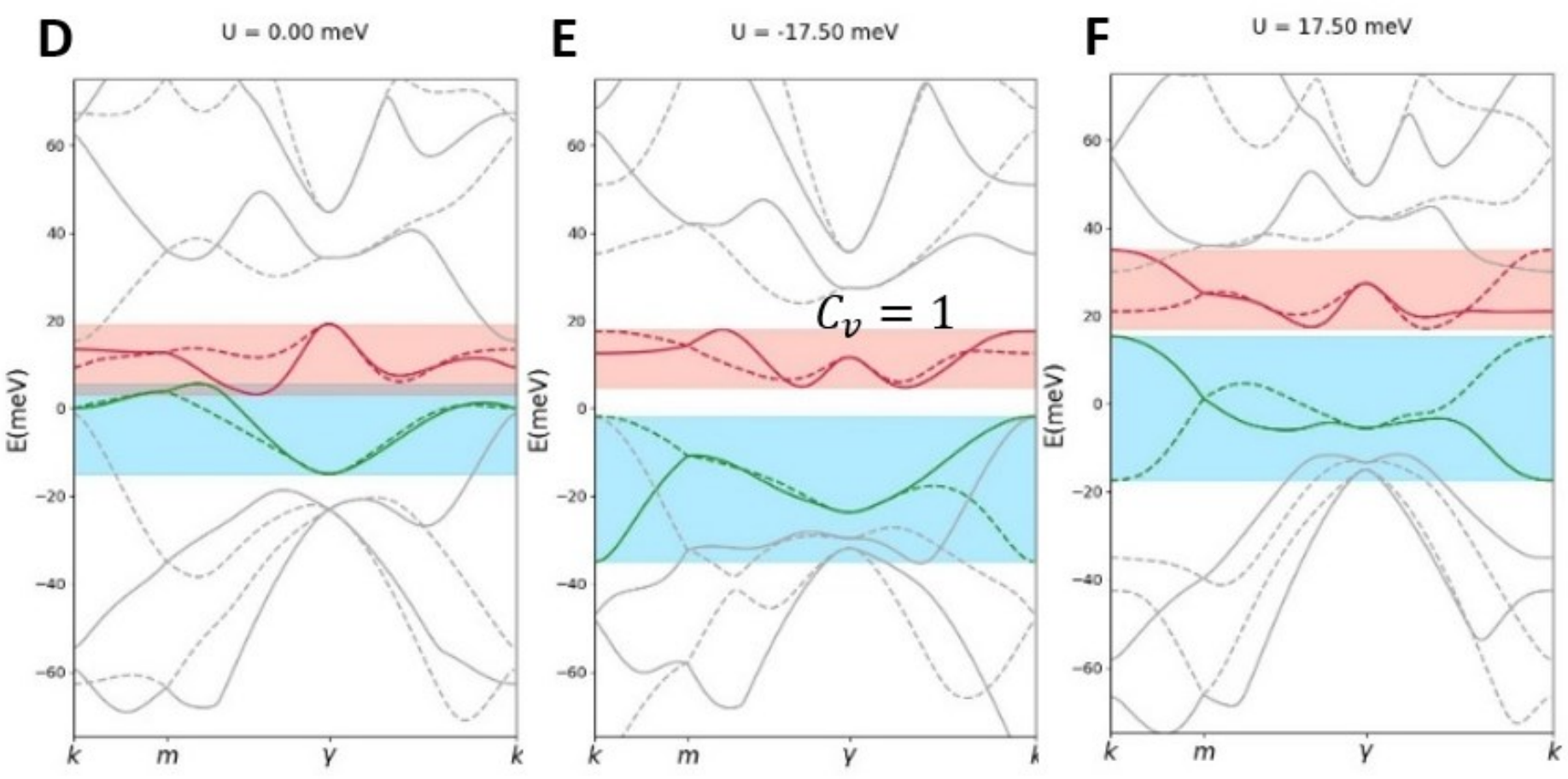

**Fig. S2.**

**Continuum model calculations. (A-F)** Band structure of 1.3° twisted rhombohedral trilayer-bilayer graphene (A-C) and twisted Bernal trilayer-bilayer graphene (D-F) at different electrical fields calculated by the continuum model. At optimal $U$, the conduction band of AB-ABA configuration exhibits a topologically isolated flat band with bandwidth ~ 15 meV and $C_v = 1$, and the conduction band of AB-ABC configuration exhibits a topologically isolated flat band with bandwidth ~ 6 meV and $C_v = 3$.

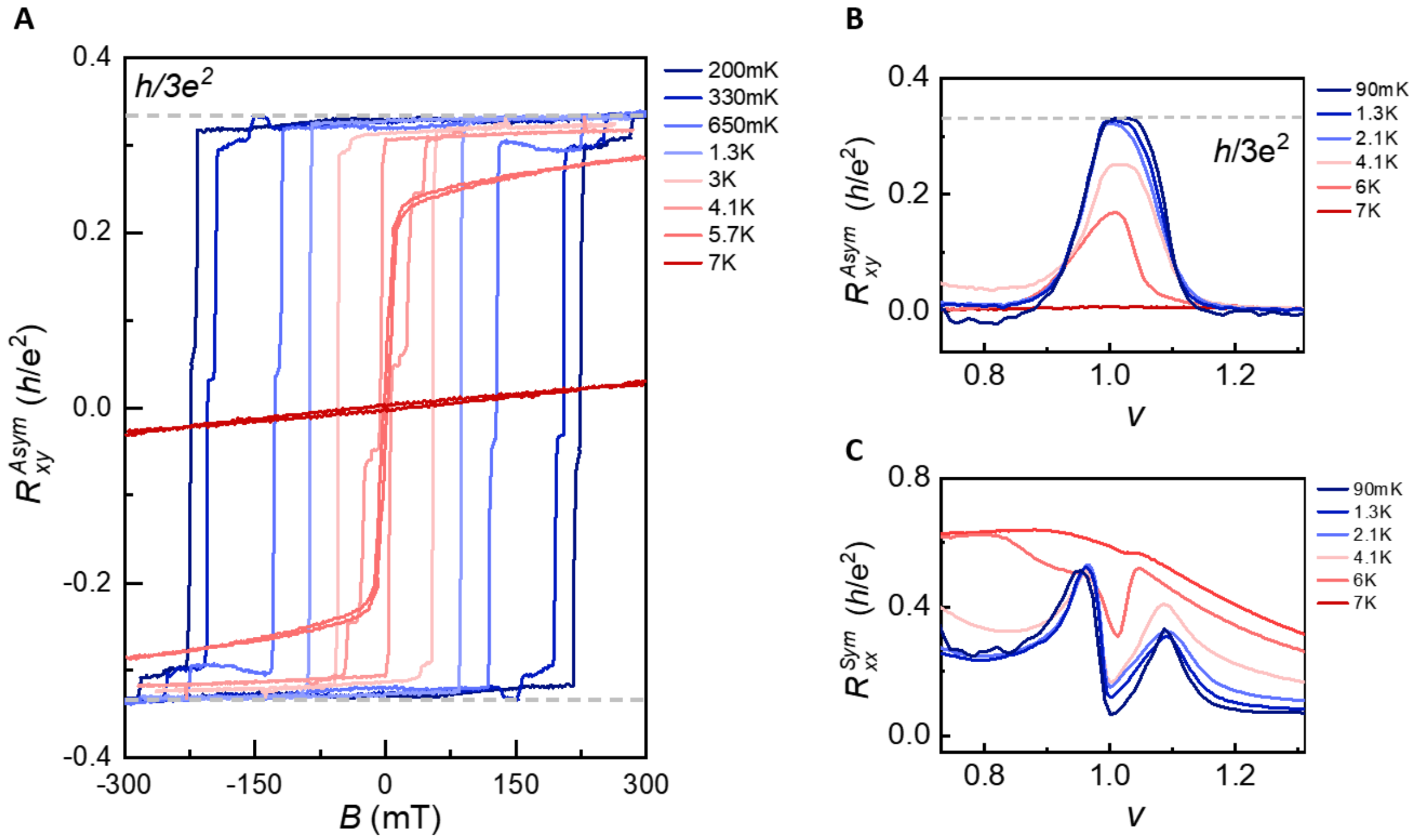

**Fig. S3.**

**Temperature dependence of the Chern insulator with C = 3 at $\nu$ = 1 in device D1.** (**A**) Quantum anomalous Hall effects of the Chern insulator at different temperatures. (**B**) Plots of anti-symmetrized $R_{xy}(\nu)$ at different temperatures under $B$ = 0.1T. (**C**) Plots of symmetrized $R_{xx}(\nu)$ at different temperatures under $B$ = 0.1T.

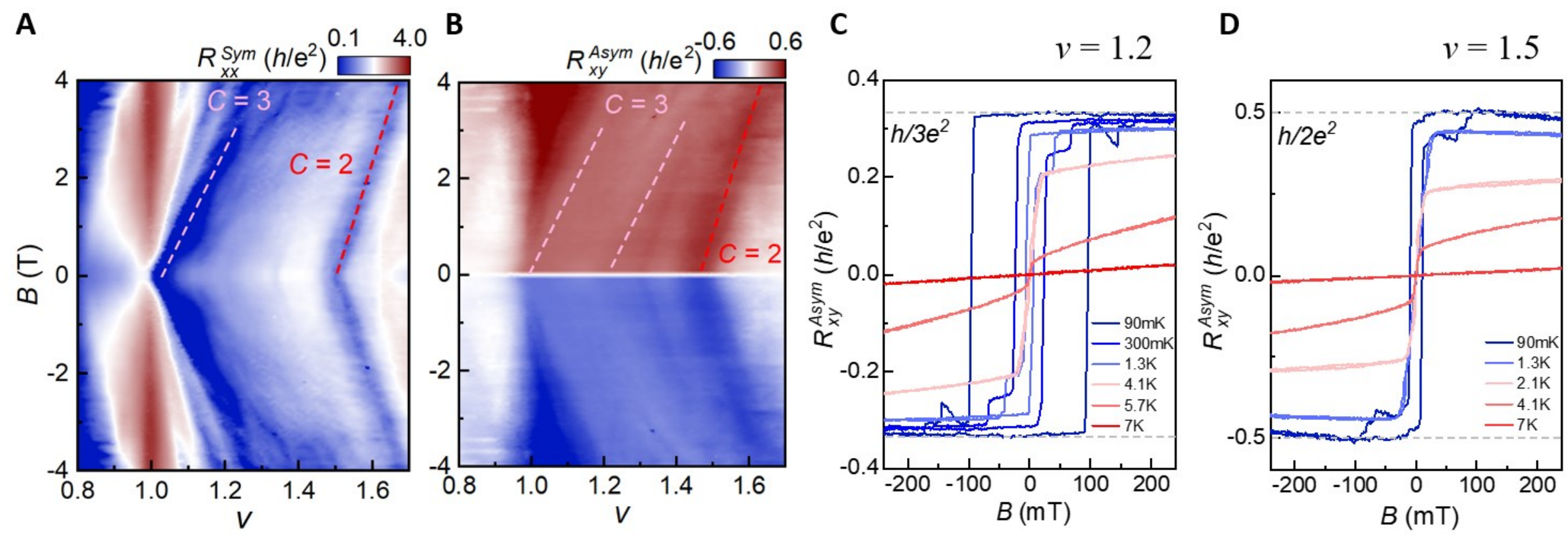

**Fig. S4.**

**Chern insulators with integer Chern numbers at fractional fillings near $\nu$ = 1 in device D1.** (**A** and **B**) Landau fan diagrams with two kinds of Chern insulators. (**C**) Temperature dependence of the QAH with $C$ = 3 at $\nu$ = 1.2 and $D$ = 0.443 V/nm. (**D**) Temperature dependence of the QAH with $C$ = 2 at $\nu$ = 1.5 and $D$ = 0.515 V/nm.

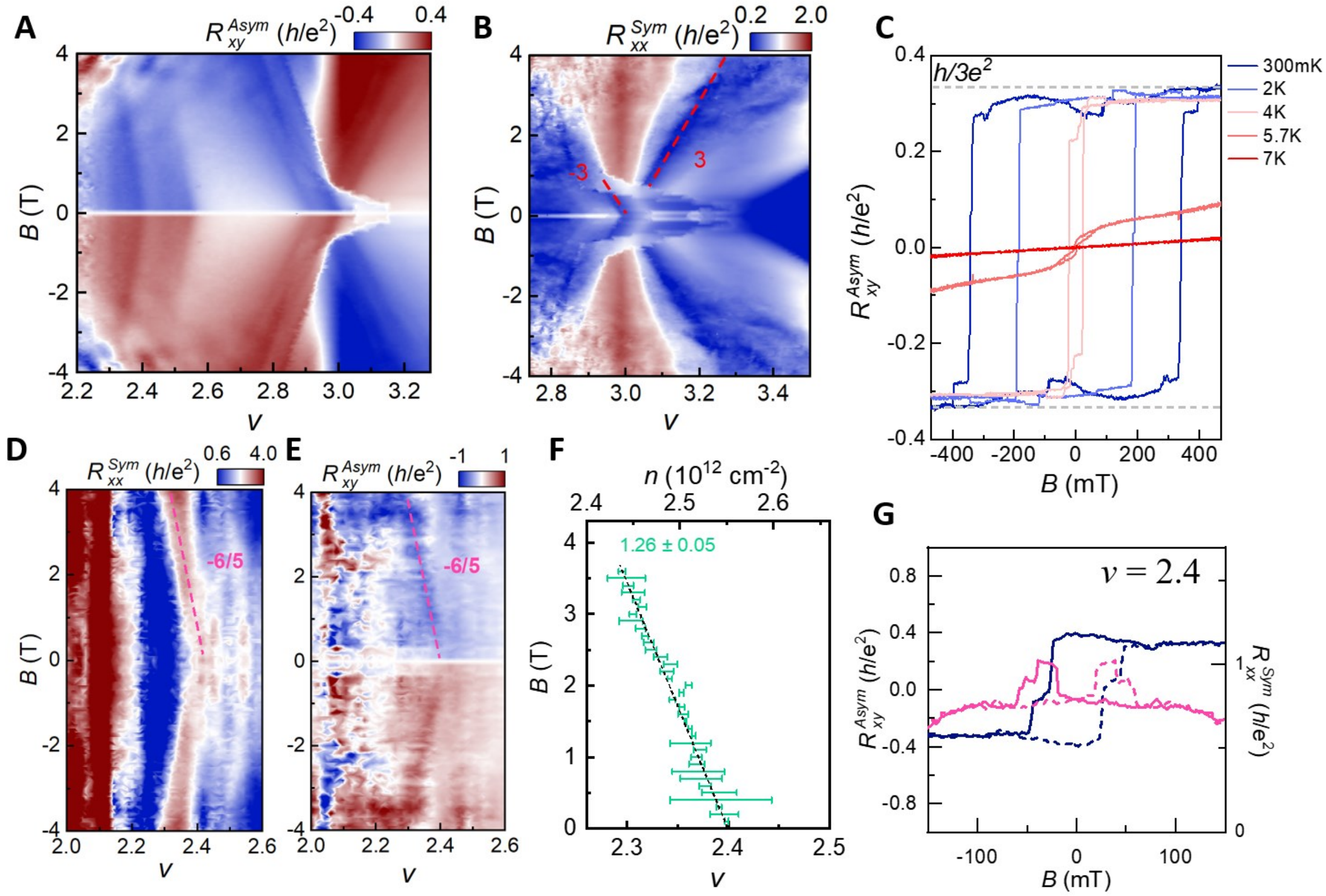

**Fig. S5.**

**Chern insulators and fractional Chern insulators near *ν* =3 in device D1.** (**A**) LaXndau fan diagram of anti-symmetrized $R_{xy}$, with integer Chern insulator at $\nu = 3$ and signature of FCI at fractional fillings. Data is taken along the line at $D = 0.35$ V/nm in Fig. 1c and $T$ =1K. Note that it is difficult to extract Chern numbers at fractional fillings at this temperature, especially for the state near v=5/2. (**B**) Landau fan diagram of symmetrized $R_{xx}$ at $D = 0.35$ V/nm and $T$ =1K, revealing ground state with $C$ = -3 at small magnetic fields and a competing one with $C$ = 3 at high magnetic fields. The observation indicates a competition between opposite valley polarization near $\nu = 3$. (**C**)Temperature dependence of the QAH with $C = 3$ at $\nu = 3$. (**D** and **E**) Landau fan diagrams to demonstrate the FCI state at $\nu$ =2.4 at $T = 0.1$K. (**F**) Extraction of the Chern number of the FCI by Streda's formula analysis from (D) and (E). (**G**) Anomalous Hall effects of the FCI at $\nu = 2.4$ and $D = 0.347$ V/nm.

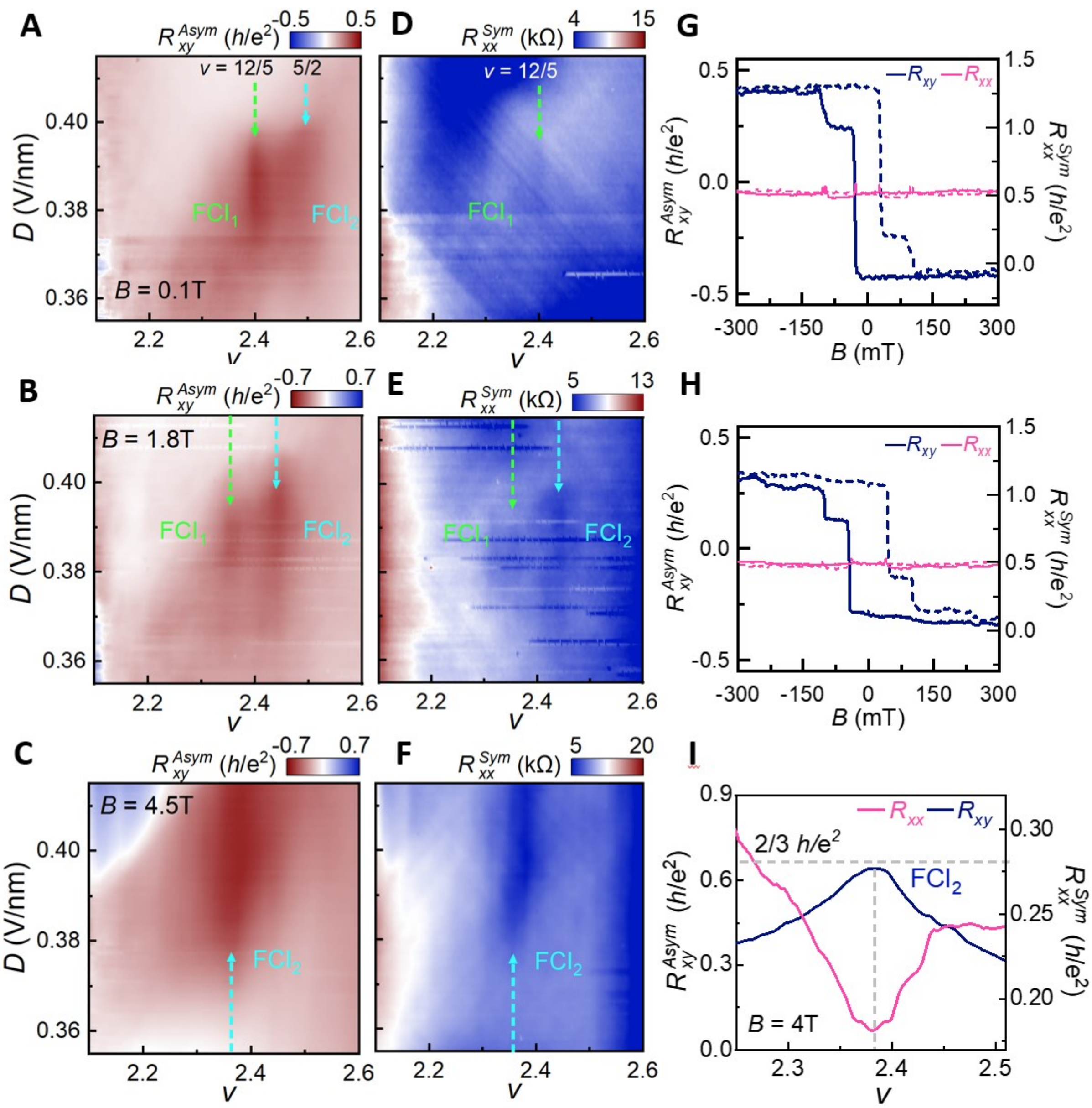

**Fig. S6.**

**Phase diagram of FCI₁ and FCI₂ States at infinite magnetic field in device D2.** (**A-C**) Zoom-in phase diagram of anti-symmetrized $R_{xy}$ at $B$ = 0.1 T (A), 1.8T (B) and 4.5T (C) as a function of filling factor and displacement field. (**D-F**) Zoom-in phase diagram of symmetrized $R_{xx}$ at $B$ = 0.1 T (D), 1.8T (E) and 4.5T (F) as a function of filling factor and displacement field. $FCI_1$ and $FCI_2$ states are indicated by green and blue dashed lines and arrows, respectively. (**G** and **H**) Anomalous Hall effect at fractional fillings $v$ = 12/5 with $D$ = 0.384 V/nm (G) and at $v$ = 5/2 with $D$ = 0.391 V/nm (H). (**I**) Plots of $R_{xy}$ (blue) and $R_{xx}$ (red) as a function of $v$ to show the $FCI_2$ at $D$ = 0.384 V/nm and $B$ = 4T. Data are measured at $T$ = 65mK.

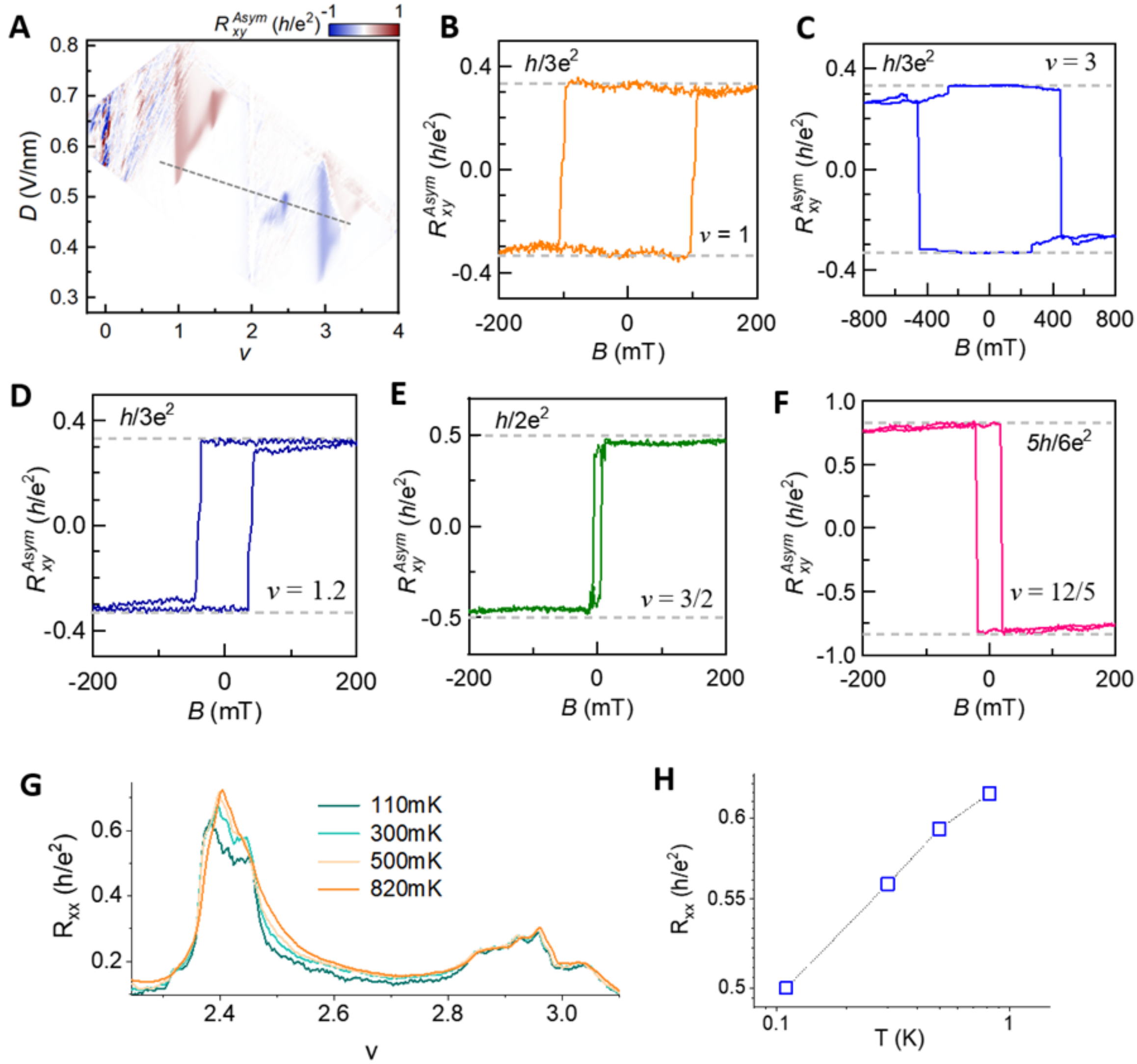

**Fig. S7.**

**Phase diagram, QAH with different Chern numbers, and fQAH in device D3.** (**A**) Color mappings of $R_{xy}$ as a function of displacement field $D$ and filling factor $v$ at $T = 0.5$ K in device D3 with $\theta = 1.5°$. The Landau fan diagram in Fig. 4a is measured along the dashed line. (**B-F**) Plots to show QAH with $C = 3$ at $v = 1$ and $D = 0.57$ V/nm (**B**), $C = -3$ at $v = 3$ and $D = 0.425$ V/nm (**C**), $C = 3$ at $v = 1.2$ and $D = 0.61$ V/nm (**D**), $C = 2$ at $v = 3/2$ and $D = 0.676$ V/nm (E) and $C = -6/5$ at $v = 2.43$ and $D = 0.494$ V/nm (F). (**G**) Plots of $R_{xx}(v)$ at different temperatures, revealing a weak dip near v = ~12/5. (**H**) Temperature dependence of $R_{xx}$ at v = ~12/5.

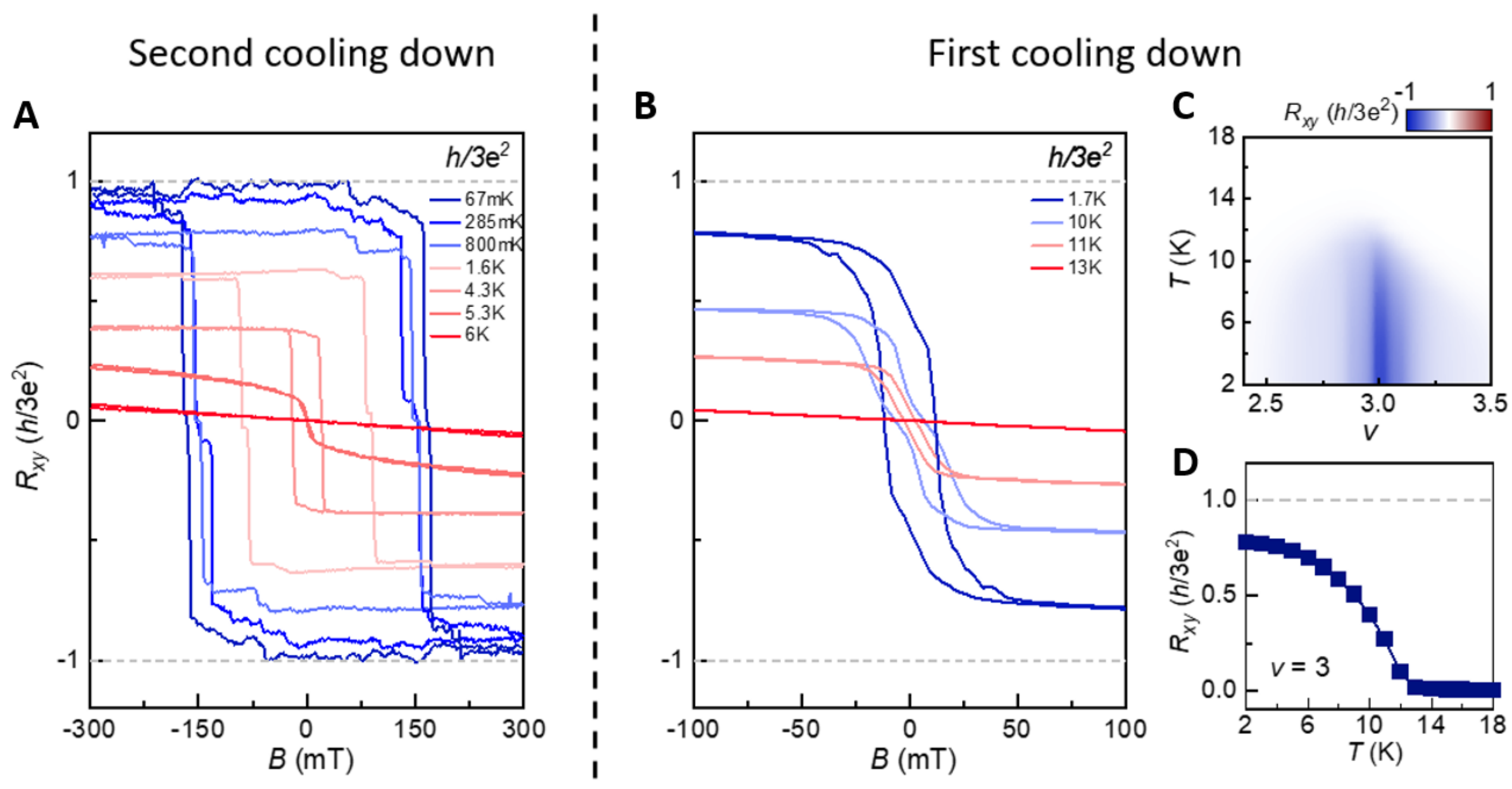

**Fig. S8.**

**Temperature dependence of the quantum anomalous Hall effect at $1^{st}$ and $2^{nd}$ cooling down in device D2.** (**A** and **B**) Magnetic hysteresis loop of $R_{xy}$ and $R_{xx}$ at $v = 3$ and $D = 0.408$ V/nm and $T = 65$mK-6K in $2^{nd}$ cooling down (A) and at $v = 3$ and $D = 0.4$ V/nm and $T = 1.7$K-13K in $1^{st}$ cooling down (B). (**C**) Anti-symmetrized $R_{xy}$ as a function of $T$ and $v$ at $B = 0.1$T in $1^{st}$ cooling down. (**D**) Temperature dependence of anti-symmetrized $R_{xy}$ at $v = 3$ and $D = 0.4$ V/nm and $B = 0.1$T. Critical temperature $T_c$ is around 13K in $1^{st}$ cooling down.

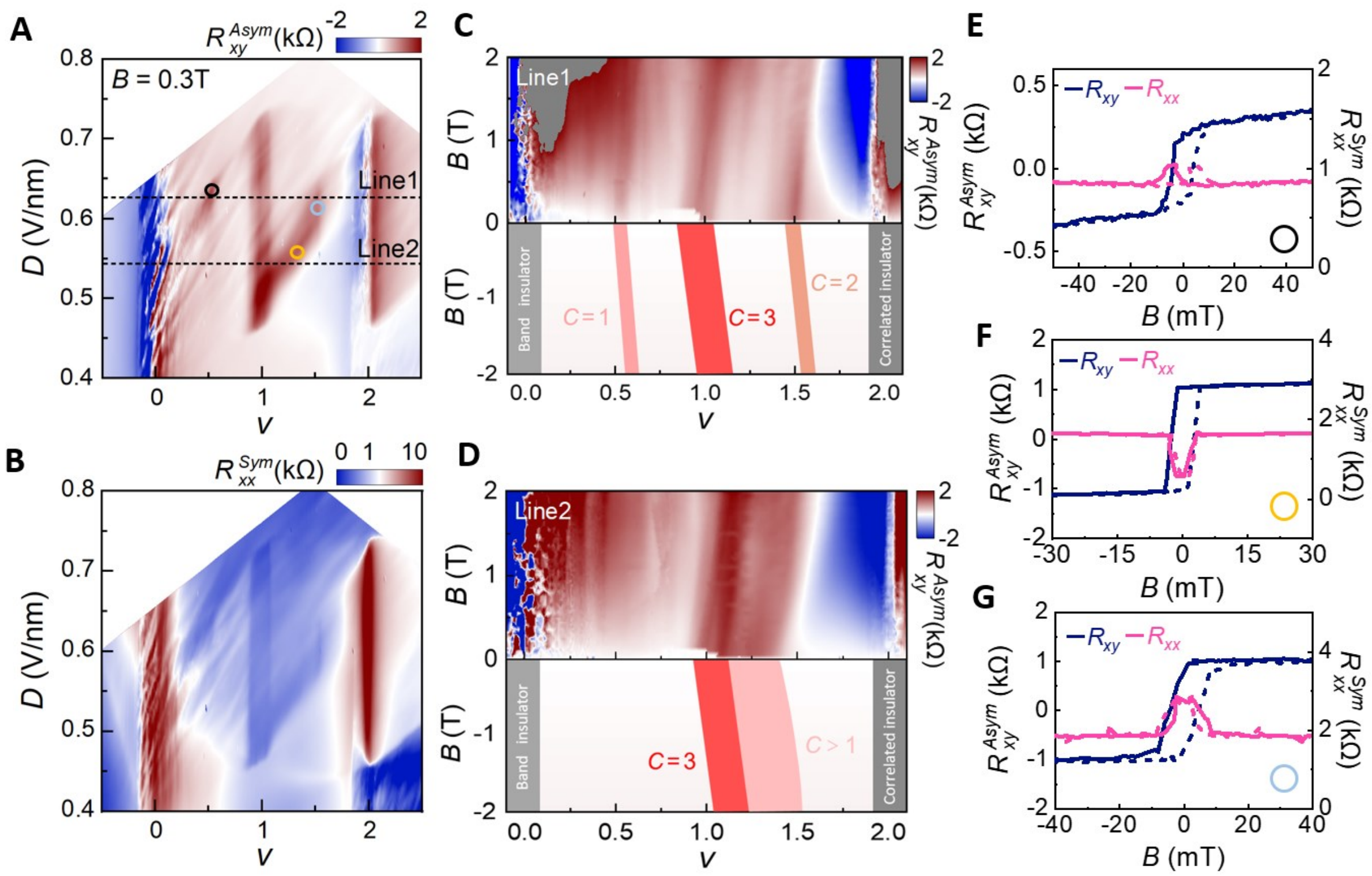

**Fig. S9.**

**Evidence of Chern insulators with C =1, 2, and 3 in device D4.** (**A** and **B**) Phase diagram of anti-symmetrized Hall resistance $R_{xy}$ (A) and $R_{xx}$ (B) at $B$ = 0.3T and $T$ = 1K. (**C** and **D**) Landau fan diagram measured along the black dashed lines in **a**. $R_{xy}$ (upper panel) measured at $D$ = 0.62 V/nm (Line1, C) and at D = 0.546 V/nm (Line2, D). The topological Chern numbers determined by Streda's formula are shown in the schematic diagram (lower panel). (**E-G**) Anomalous Hall effects measured at $\nu$ = 1/2 with $D$ = 0.637 V/nm (E), $\nu$ = 1.3 with $D$ = 0.552 V/nm (F), and $\nu$ = 3/2 with $D$ = 0.61 V/nm (G), corresponding to the three points in (A), marked by black, yellow and blue circle, respectively.

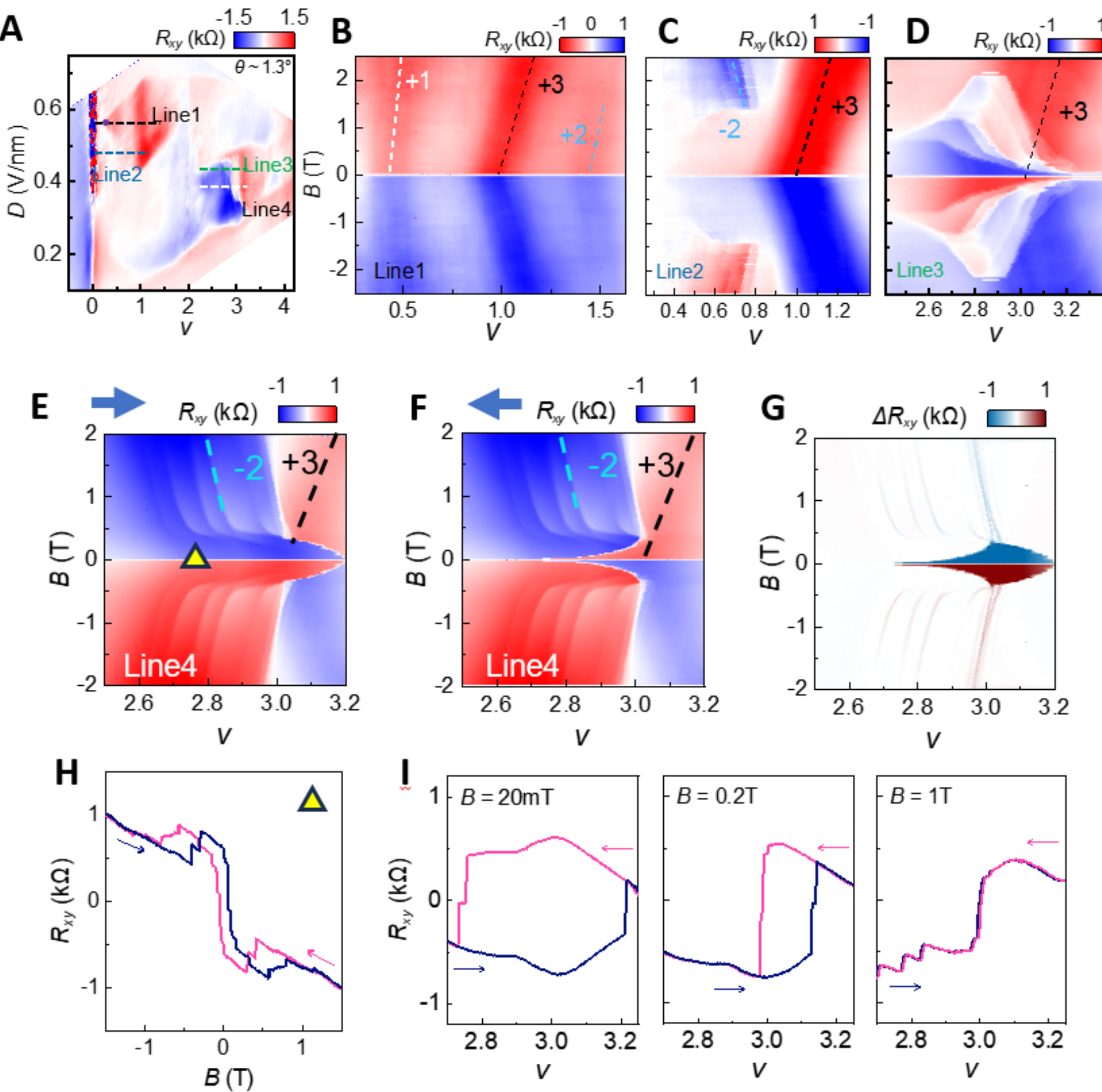

**Fig. S10.**

**Evidence of various Chern insulators with different Chern numbers and first-order phase transitions in device D5.** (**A**) Phase diagrams of anti-symmetrized $R_{xy}$ as a function of filling factor and displacement field at $B$ = 0.5 T and $T$ = 1.1 K. (**B-D**) Landau fan diagrams measured at $D$ = 0.575 V/nm (B), $D$ = 0.475 V/nm (C) and $D$ = 0.446 V/nm (D) corresponding to dashed line1, line2, and line3 in (A), respectively. The slopes of white, blue, and black dashed lines correspond to $C$ = 1, $\pm$2, and 3. (**E** and **F**) Landau fan diagram of $R_{xy}$ at $D$ = 0.385 V/nm with different scanning directions, corresponding to white dashed line 4 in (A). The blue arrows at the top of the figure, pointing to the right and left, represent scanning forth (E) and back (F), respectively. The black dashed line indicates the Chern insulator with $C$ = 3, and the blue dashed line indicates topological phases with $C$ = -2 determined by Streda's formula. (**G**) Electric-field hysteresis map extracted from (E) and (F). The hysteresis loop disappears at $B$ > 0.4 T. At each fixed $B$, $\Delta R_{xy}$ is obtained by subtracting the Hall resistance in (E) from that in (F). **(H)** Magnetic hysteresis of $R_{xy}$ at $\nu$ = 2.75 and $D$ = 0.385 V/nm, corresponding to the yellow triangle in (E). (**I**) Electric-field hysteresis loop at different magnetic field.

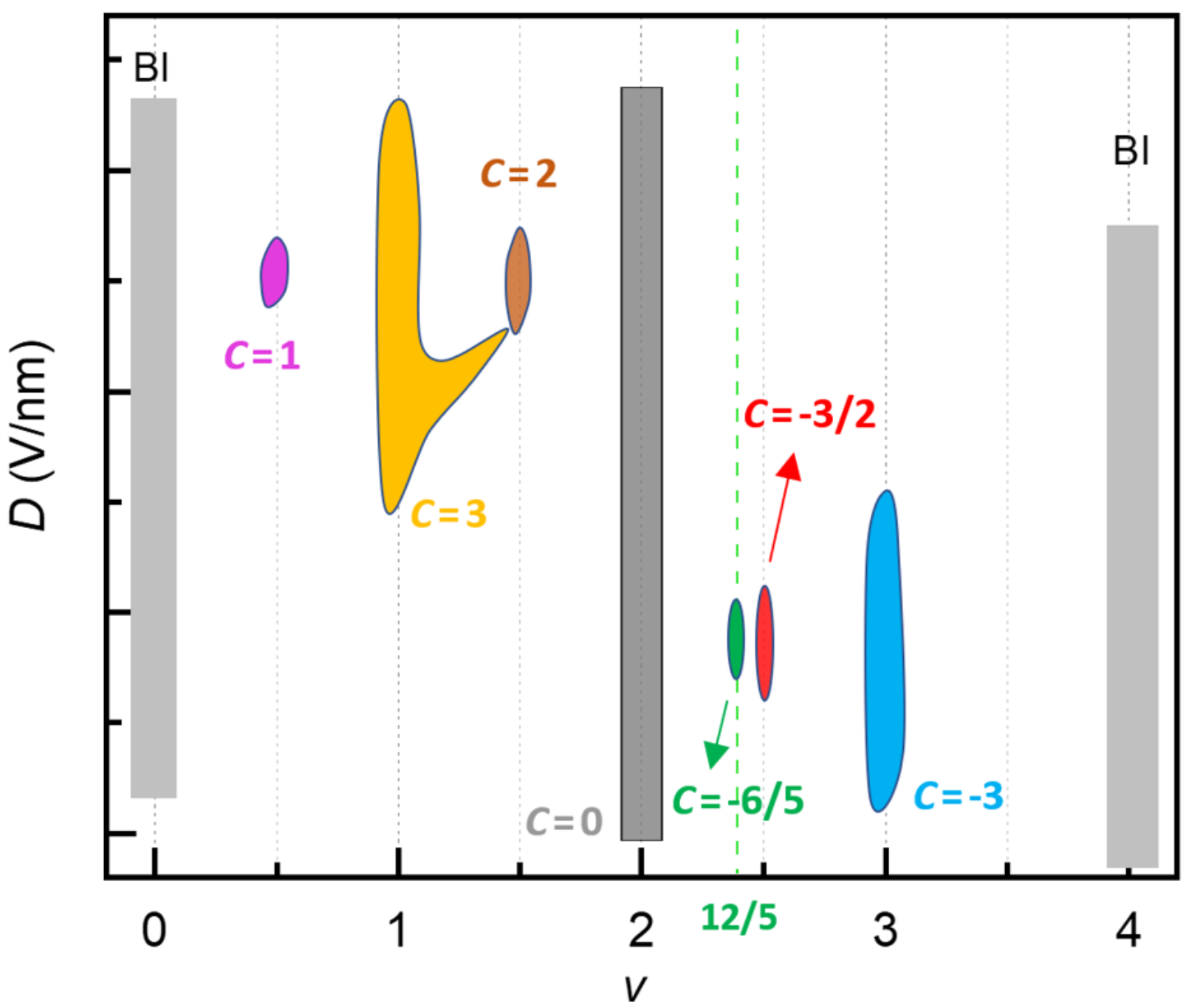

**Fig. S11.**
**A Schematic phase diagram of integer Chern insulators and fractional Chern insulators observed in tRTBG.**